\documentclass[aps,pre,epsfigure,twocolumn,superscriptaddress,nobibnotes]{revtex4-2}
\usepackage{dcolumn} 
\usepackage{bm} 
\usepackage{graphicx}
\usepackage{amsmath}    
\usepackage{latexsym}
\usepackage{amsfonts}   
\usepackage{bbold}
\usepackage{amssymb}
\usepackage{breqn}
\usepackage{array}      
\usepackage{epsfig}
\usepackage{txfonts}
\usepackage{color}
\usepackage{diagbox}
\usepackage[colorlinks=true,linkcolor=blue,urlcolor=blue,citecolor=blue]{hyperref}
\usepackage{hyperref}
\usepackage{enumitem}
\usepackage{bbold}
\usepackage{cleveref}
\usepackage{appendix}
\usepackage{siunitx}
\usepackage{braket}
\DeclareMathOperator{\Tr}{Tr}
\newcommand{\specialcell}[2][c]{%
	\begin{tabular}[#1]{@{}c@{}}#2\end{tabular}}

\newcommand{\hhu}{Institut f\"ur Theoretische Physik II: Weiche Materie, Heinrich-Heine-Universit\"at D\"usseldorf, Universit\"atsstr. 1, D-40225 D\"usseldorf,
	Germany}
 \newcommand{\tud}{Technische Universit\"at Darmstadt, Hochschulstra{\ss}e 8, D-64289 Darmstadt, Germany}
\begin{document}
	\title{Engineering active motion in quantum matter}
	\author{Alexander P.\ Antonov}
	\thanks{These authors contributed equally to this work}
	\affiliation{\hhu}
	\author{Yuanjian Zheng} 	
	\thanks{These authors contributed equally to this work}
	\affiliation{\hhu}

	\author{Benno Liebchen}
	\email{benno.liebchen@pkm.tu-darmstadt.de}
	\affiliation{\tud}
	\author{Hartmut L\"owen}
	\email{hlowen@hhu.de}
	\affiliation{\hhu}

	\begin{abstract}     
         We introduce a framework for engineering active quantum matter that involves mimicking the role of self-propulsion through an external trapping potential that is moving along imposed trajectories traced by classical active dynamics. This approach in the presence of dissipation, not only recovers essential dynamical behavior of classical activity, including the ballistic to diffusive cross-over of its mean-square displacement, but also reveals additional features of activity that are of quantum origin. These quantum-active features are revealed in non-dissipative systems, and manifest as novel exponents of the mean-square displacement at short timescales.
         
	\end{abstract}	
	\maketitle
	
	\section{Introduction}
 
Over the past two decades, non-equilibrium physics has been shaped by two remarkable developments: the rise of active matter, which has unveiled a plethora of out of equilibrium self-organization phenomena, and the advancement of methods enabling the precise control and manipulation of individual inherently quantum objects.

The first development -- active matter, concerns the description of agents that are capable of using energy from their environment to propel themselves \cite{MarchettiSimha2013,Ramaswamy2010,Gompper2020} -- either autonomously \cite{paxton2004catalytic, howse2007self, BechingerVolpe2016} -- or through the actuation with time-dependent external fields \cite{DreyfusBibette2005, TiernoSagues2008, DeseigneChate2010, LimJaeger2019}. Such agents, called active particles were first realized at the micron-scale about two decades ago \cite{paxton2004catalytic}, and induce a spectacular plethora of non-equilibrium phenomena. For a single particle, this includes the emergence of a mean squared displacement (MSD) beyond the generic Stokes-Einstein relation and characteristic ballistic motion that enables the usage of active particles as efficient micro-motors for targeted drug delivery. At the many-body level, the local driving of active systems opens up a world of spectacular phenomena. This includes the spontaneous emergence of a center of mass motion induced by non-reciprocal interactions of isotropic particles \cite{soto2014self,ivlev2015statistical,niu2018dynamics,schmidt2019light,saha2020scalar,grauer2021active,SurowkaBanerjee2023}, and the emergence of novel types of phase transitions that lead to so-called flocks where particles align over large distances and move collectively in a spontaneously chosen direction. These flocks exploit non-equilibrium fluctuations to bypass the Mermin-Wagner theorem and exhibit true long-range order in two-dimensions \cite{toner1995long, toner1998flocks}, showing giant number fluctuations that violate the central limit theorem, making it impossible to measure (predict) the density in a finite domain \cite{narayan2007long}. Other phenomena specific to active matter include motility-induced phase separation where an ensemble of purely repulsively interacting particles spontaneously phase separate into a co-existing gas-like and a liquid-like phase \cite{cates2015motility}, as well as living clusters that dynamically form, break-up and reform recursively \cite{theurkauff2012dynamic,palacci2013living,buttinoni2013dynamical,ginot2018aggregation}. Many of these phenomena have been show in the past decade to occur at a broad range of scales, ranging from macroscopic robots, birds and drones \cite{Whitesides2018, LeymanVolpe2018, VasarhelyiVicsek2018, AraujoVolpe2023}, down to the micro- and larger nanoscale, i.e.,~10  \si{\nano\meter}--1 \si{\micro\meter}, where bacteria, molecular motors and synthetic colloidal microswimmers thrive  \cite{Ebbens2016, ShaebaniRieger2020, MognettiFrenkel2013, Klapp2016, LeunissenBlaaderen2005, ZiepkeFrey2022}. 

The second development revolves around the evolution of cold atom physics, that was invigorated by the observation of Bose-Einstein condensation in dilute atomic gases \cite{BEC1,BEC2}. Ever since, much progress has been achieved regarding the developments of techniques for cooling, trapping and the controlled manipulation of quantum objects both at the single \cite{hu1994observation,leibfried2003quantum} and the many-particle level \cite{lewenstein2007ultracold,bloch2008many}, even in non-equilibrium situations \cite{polkovnikov2011colloquium,langen2015ultracold,schneider2012fermionic,heyl2018dynamical}.

This co-discovery of new phenomena in active matter and novel methods for controlling atoms in the deep quantum regime, begs the question of whether novel active non-equilibrium phase transitions and pattern formation could also be realized at a scale where quantum effects must be considered. In particular, one may wonder if there is a quantum-analog to active phenomena like motility-induced phase separation, dynamic clustering or giant density fluctuations. However, progress in understanding this interplay has been limited thus far, since it is yet unknown how a self-propelled particle could be realized in the quantum regime. Accordingly, existing attempts to investigate active systems at the quantum scale are limited to a relatively small number of pioneering works either in identifying and exploring original 
formal analogies between descriptions of quantum systems and classical active behavior \cite{Vicsek1995,LopezClenebtt2015, GuoCheng2018, teVrugtWittkowski2023, SoneAshida2019,LoeweGoldbart2018,CouderFort2006}, or examining innovative models involving classical swimmers embedded in quantum media such as fermionic liquids \cite{AvronOaknin2006}, Bose-Einstein condensates \cite{Saito2015} and super-fluid He$^4$ \cite{ShuklaPandit2016, GiuriatoKrstulovic2019, KolmakovAranson2021}. In all these semi-classical approaches, the driven agent remains a classical object. Recently, the role of activity \cite{AdachiKawaguchi2022, YamagishiObuse2023} in the form of hopping terms across lattice sites, has also been considered in quantum many-body systems with non-hermitian Hamiltonians \cite{MatsoukasDelcampo2023,quantumflocks}. These highly original approaches allow exploration of activated dynamics on a lattice, which does however restrict them from exploring generic active matter phenomena in continuum space. 

\begin{figure}[t!]	
\includegraphics[width=\columnwidth]{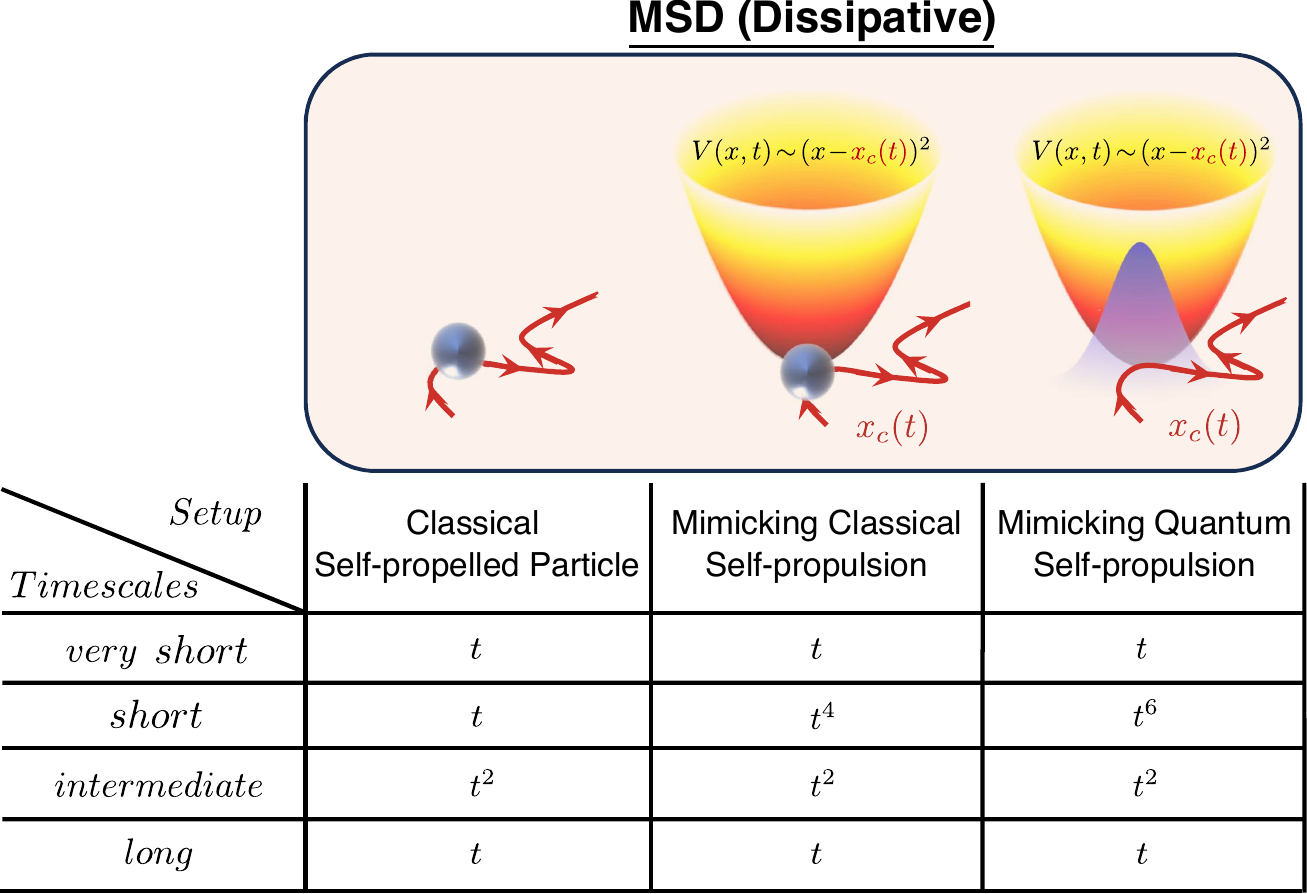}		\caption{[{\bf  Schematics and dynamics of dissipative quantum and classical overdamped active systems}]. [Left] Classical self-propelled particle (blue) with trajectory (red) and corresponding characteristic scaling of the MSD of an ensemble of self-propelled particles. 
[Center] Displacing a passive classical particle with a moving harmonic trapping potential (red to yellow paraboloid) mimics self-propulsion. [Right] Replacing the trapped classical particle by a quantum particle (wavefunction in blue) enables mimicking of quantum self-propulsion.
}
	\label{fig:schematic_dissipative}
	\vspace{-2.5ex}
\end{figure}

In this work, we present a framework that is orthogonal to these existing approaches. This allows us to \emph{bypass the challenge of realizing intrinsically self-propelled quantum particles}. The central idea is to \emph{mimic self-propulsion} by guiding quantum particles with a moving external trapping potential along the path of an active particle (Fig.~\ref{fig:schematic_dissipative}). We show that this driven quantum particle, despite not self-propelling autonomously, recovers all desirable properties of a active quantum particle (Fig.~\ref{fig:schematic_dissipative}). That is, our approach leads, to a fully quantized Hamiltonian that describes a quantum particle that: 
\setlist{nolistsep}
\begin{enumerate}[label=\roman*., noitemsep]
\item breaks local detailed balance
\item moves in continuum space
\item leads to a MSD that is quadratic at intermediate timescales in the presence of dissipation \cite{BechingerVolpe2016}.
\item leads to a MSD that is linear at long timescales (active diffusion \cite{BechingerVolpe2016}). 
\end{enumerate}

In particular, this approach is beyond semi-classical and lattice-based realizations of active matter and leads to new active quantum effects even at the single particle level that are accentuated in the absence of dissipation. 

While the focus of this work is on introducing a  minimal framework of active quantum particles, and accordingly, exploration of collective phenomena is beyond its scope, we emphasize that this approach allows for an almost direct realization of single and collective active matter phenomena with quantum particles. Our approach can be realized using 
ultra-cold atoms in time-dependent optical traps \cite{BlochNascimbene2012, Kaufman2021, Lahaye2020, SchaferTakahashi2020,MorigiWineland2015}, in which the positions of the potential minimum can be externally programmed almost at will and at very high speeds \cite{PesceVolpe2020, Ashkin1970, ChenMuga2010, AbahLutz2012, RossnagelLutz2014}. Our work thus opens a door to a new class of materials, namely that of active quantum matter.

\section{Mimicking quantum propulsion}

We now consider a quantum particle trapped in a time-dependent potential which is uniquely parameterized by its minimum position $x_c(t)$.\ This position is externally programmed to follow random paths drawn from an ensemble of trajectories generated by classical active dynamics \cite{SabhapanditMajumdar2024, UhlenbeckOrnstein1930, BechingerVolpe2016, RomanczukGeier2012}.  Specifically, we consider active Ornstein-Uhlenbeck particles (AOUP), where the velocities follow a classical stochastic process with both a random diffusive component controlled by the diffusion constant $D$ and persistent motion governed by the persistence time $\tau$ \cite{Bonilla2019}, given by the equation of motion:
\begin{equation}
\tau \ddot{x}_c= -\dot{x}_c+\sqrt{2D}\eta(t)
\label{eq:active_eom}
\end{equation}
where $\eta(t)$ is a zero-mean delta-correlated white noise. We now write the time-dependent Hamiltonian as
\begin{equation}
	\hat{H}(t)=\frac{\hat{p}^2}{2m} + \hat{V}(\hat{x}, t),
	\label{eq:hamiltonian}
\end{equation}
where $\hat{V}(\hat{x},t)$ is the potential energy operator associated with a classical trapping potential $V(x,t)$;
$\hat{p}=-i\hbar\partial_{x}$ is the momentum operator, and $m$, $\omega$, $\hat{x}$ are the particle mass, harmonic trap frequency and position operator respectively. In the following, we consider two cases, dissipative and non-dissipative. In the \textit{dissipative} case, we assume that the quantum particle is weakly coupled to its environment such that the dynamics of its density matrix $\hat{\rho}(t)$ is governed by the Lindblad equation \cite{RivasHuelga2011}:
\begin{equation}
\dot{\hat{\rho}}=-\frac{i}{\hbar}\left[ \hat{H}(t),\hat{\rho}\right]+\hat{\mathcal{D}}(\hat{\rho},t)
\label{eq:time_evolution}
\end{equation}
and the (Markovian) dissipator is given by 
\begin{equation}
\hat{\mathcal{D}}(\hat{\rho},t)=\nu_{+} \left(  \hat{a}^\dagger \hat{\rho} \hat{a} - \frac{1}{2}\left\{\hat{a}\hat{a}^\dagger ,\hat{\rho} \right\} \right) + \nu_{-} \left(  \hat{a} \hat{\rho} \hat{a}^\dagger - \frac{1}{2} \left\{\hat{a}^\dagger \hat{a} ,\hat{\rho} \right\} \right) 
\label{eq:dissipator}
\end{equation}
where $\hat{a} $ ($\hat{a}^\dagger$) is the annihilation (creation) operator, while $\nu^{+}$ and $\nu^{-}$ are the gain and loss rates for the excitations of the quantum particle in the trap respectively, and correspond to an equilibrium temperature ($T$) of the thermal bath through $\nu_{+}/\nu_{-}=e^{-\hbar \omega /k_B T}$ where $k_B$ is the Boltzmann constant and $\hbar$ is the reduced Planck constant \cite{EnglertMorigi2002}. In the \textit{non-dissipative} (or Hamiltonian) case, the dissipator in Eq.~\eqref{eq:time_evolution} vanishes, $\hat{\mathcal{D}} \equiv 0$. We note that the used form of the Lindblad equation is one specific choice to model the coupling to a thermal bath. In a future work, we will examine how the predicted quantum-active effects depend on the specific choice of the bath-coupling model \cite{Lindbladcorr}.\\
\indent We define the quantum MSD, or the spread of the time-dependent wavefunction $|\psi_t\rangle$, as the expectation value: 
\begin{eqnarray}
\text{MSD} & \equiv &\overline{\langle \psi_t |\hat{x}^2|\psi_t \rangle} - \langle \psi_{t=0} |\hat{x}^2|\psi_{t=0} \rangle \nonumber \\
& = & \overline{\textrm{Tr}({\hat{\rho}(t)\hat{x}^2})} - \textrm{Tr}({\hat{\rho}(0)\hat{x}^2}).
\label{eq:msd-define}
\end{eqnarray}
Here, $\overline{\vspace{2ex}\cdots}$ denotes the average over prescribed stochastic trajectories, that we will specify below. In what follows, we analyze equation \eqref{eq:time_evolution} both analytically and numerically, starting from the ground state $|\psi_{t=0}\rangle$, which is the quantum analogue
	of initialization in the potential minimum in the classical case. Further details of our analysis and methods provided in the Supplemental Material (SM) \cite{supplement}. We choose $\sqrt{\hbar/m\omega}$ and $\omega^{-1}$ as units of length and time, respectively.

\begin{figure}[t!]	
\includegraphics[width=\columnwidth]{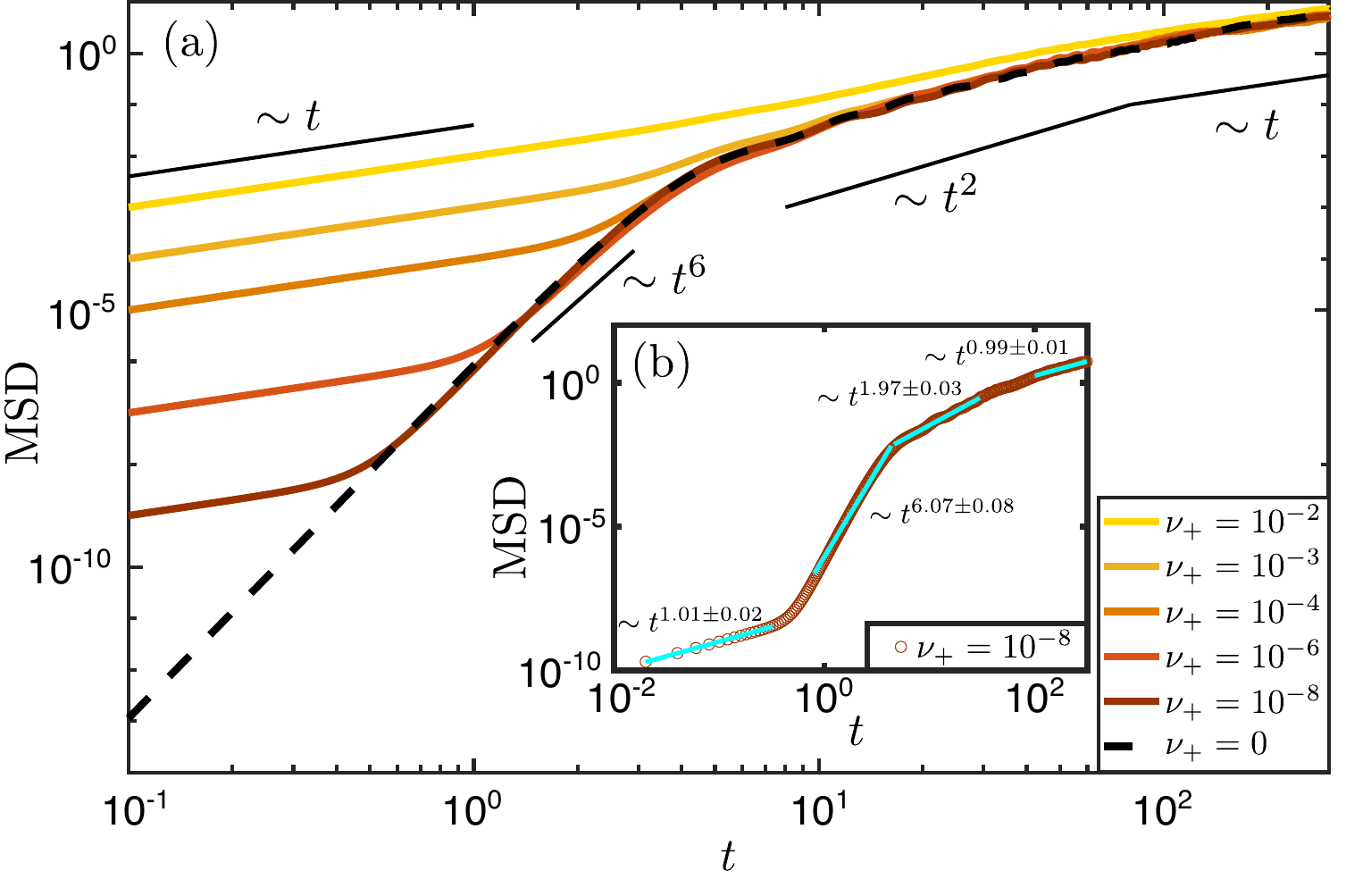}		
	\caption{{[\bf MSD of a dissipative quantum-active particle]}. (a) MSD at various timescales for $D=0.01$, $\tau=10$ and $\nu_{-}=10^{-2}$. (b) Fits of the MSD (cyan solid lines) for $\nu_{+}=10^{-8}$ indicating the scaling behavior presented in Fig.~\ref{fig:schematic_dissipative} for a quantum-active particle. Uncertainties represent the standard deviation across $500$ simulations and do not account for possible systematic numerical uncertainties.}
	\label{fig:msd_dissipative}
\end{figure}

\section{Dissipative case} 

Here, we consider a harmonic trapping potential,
\begin{equation}
    V(x,t) = \frac{1}{2}m\omega^2(x-x_c(t))^2,
    \label{eq:harmonic}
\end{equation}
where $x_c(t)$ is a stochastic path, that is individually generated for each initial condition by solving Eq.~\eqref{eq:active_eom}.\ Figure~\ref{fig:schematic_dissipative} illustrates the scenarios that we discuss in the following: a classical self-propelled particle (left), a passive particle in the potential $V(x,t)$ (center), and a quantum particle with a corresponding potential energy operator $\hat{V}(\hat{x},t)$ (right).
Numerical results for the quantum MSD scaling are shown in Fig.~\ref{fig:msd_dissipative}, while the analytical calculation of the dynamical exponents in both classical cases is provided in the SM.

At very short timescales ($t\!\lll\!\tau$, first row in Fig.~\ref{fig:schematic_dissipative}), the quantum MSD scales diffusively ($\rm{MSD}\!\sim\!t$), reflecting the behavior observed in both classical cases. In Fig.~\ref{fig:msd_dissipative}, we identify this diffusive regime for any non-zero temperature (i.e.,\ $\nu_+\!\ne \!0$), which is also confirmed analytically in the SM, Sec.~\ref{sec:dissipation}. However, while classical active particle demonstrate this diffusive behavior also for short timescales ($t\! \ll \! \tau$, second row in Fig.~\ref{fig:schematic_dissipative}), displacing a particle by a moving trapping potential leads to a distinct regime for both classical and quantum particles mimicking self-propulsion. This regime is characterized by a rapid growth of the MSD, scaling as $t^6$ in the quantum-active case (Fig.~\ref{fig:msd_dissipative}(a)) and as $t^4$ in its classical counterpart. We note that this quantum $t^6$-scaling is also not observed
	in the underdamped classical case \cite{nguyen2021active}. The duration of this regime for the quantum particle increases as the temperature $T$ decreases (i.e.,\ for lower $\nu_+$), due to a progressive shortening of the initial diffusive regime. This suggests that the anomalous regime has a quantum origin, as quantum effects arising from the Hamiltonian dynamics become more pronounced at lower temperatures. In fact, importantly, at zero temperature (i.e.,\ $\nu_+=0$), the initial diffusive regime of the MSD disappears entirely (black dashed line in Fig.~\ref{fig:msd_dissipative}(a)). At intermediate timescales ($t \! \sim \! \tau$, third row in Fig.~\ref{fig:schematic_dissipative}), the quantum-active dynamics exhibits a ballistic regime ($\textrm{MSD}  \!\sim \!t^2$), before eventually entering a diffusive regime ($\textrm{MSD} \!\sim \! t$) at much longer times ($t \!\gg \!\tau$, fourth row in Fig.~\ref{fig:schematic_dissipative}). Both these regimes are observed in corresponding classical cases. Power-law fits of the MSD in various dynamical regimes for $\nu_+=10^{-8}$ (Fig.~\ref{fig:msd_dissipative}(b)) clearly reveal all the regimes listed in Fig.~\ref{fig:schematic_dissipative}. 
Thus, a quantum particle in a harmonic trapping potential that moves along predetermined stochastic trajectories exhibits all the characteristic properties of classical activity (i-iv in the introduction) -- and additionally points to quantum-active effect occurring at short times.

\section{Non-dissipative case} 

To further explore the significance of quantum-active effects, without partially masking it with contributions from the thermal bath, we now focus solely on the non-dissipative (Hamiltonian) case. We consider a family of power-law trapping potentials $V(x,t)$, defined as
\begin{equation}
    V(x,t) = V_0|x-x_c(t)|^q, \ q>0,
\label{eq:potential}
\end{equation}
with the prefactor $V_0 = \frac{1}{2}\hbar^{1-\frac{q}{2}}m^{\frac{q}{2}}\omega^{\frac{q}{2}+1}$.
\begin{figure}[t!]	
\includegraphics[width=.9\columnwidth]{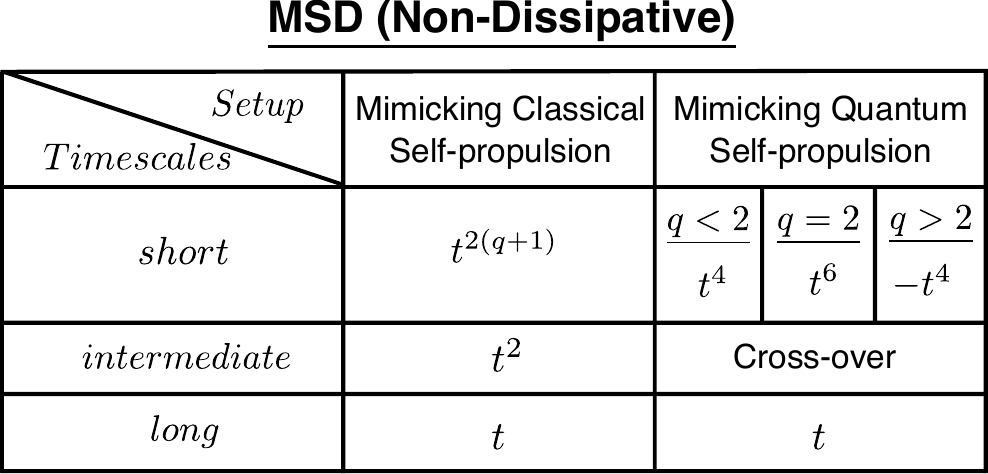}		
	\caption{[{\bf MSD of non-dissipative quantum and classical active systems}]. [Left] MSD of a classical Hamiltonian particle in a moving trapping potential $V(x,t)$ \eqref{eq:potential}; this case is akin to the center column in Fig.~\ref{fig:schematic_dissipative} but without dissipation \cite{nguyen2021active}. [Right] MSD of a Hamiltonian quantum particle in a moving trapping potential $V(x,t)$; this case is akin to the right column in Fig.~\ref{fig:schematic_dissipative} but without bath coupling. 
    }
	\label{fig:schematic_non_dissipative}
\end{figure}

The regimes emerging for classical and quantum particles mimicking self-propulsion are summarized in Fig.~\ref{fig:schematic_non_dissipative}.
Without dissipation, the diffusive regime at very short times is absent in both classical and quantum cases. We now determine the short-time ($t\!\ll\!\tau$) behavior of the MSD for a quantum particle using time-dependent perturbation theory (see the SM, Sec.~\ref{subsection:perturbation}) to obtain:
\begin{equation}
	\text{MSD} = A_q \int_0^t \int_0^t dt' dt'' \overline{x_c(t')x_c(t'')}.
	\label{eq:classical_average}
\end{equation}

\begin{figure*}[ht!]
	\center{\includegraphics[width=\linewidth]{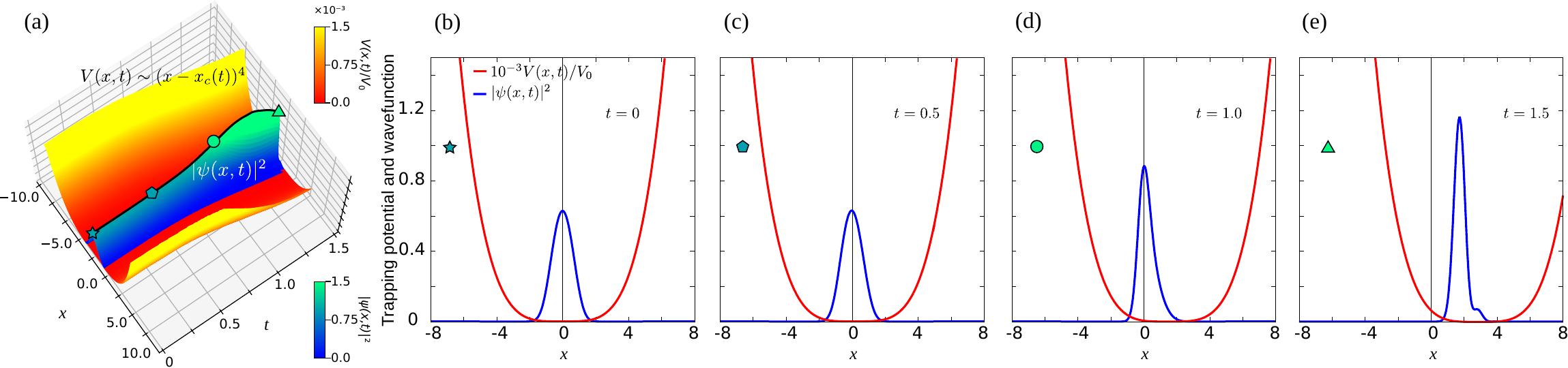}}
	\caption{[\textbf{Transient localization of the wavefunction}. (a) Time evolution of the confining potential $V(x, t)$ and the wavefunction profile $\langle \psi_t|\psi_t\rangle \equiv |\psi(x,t)|^2$, shown using red-yellow and blue-green color maps, respectively, for $D=1,\, \tau=1,\, q=4$. In the final state, the peak of the profile (shown with the black line) increases in height compared to the initial state, indicating stronger localization. (b)-(e) Potential (red) and wavefunction profiles (blue) at different time instants $t = 0,\, 0.5,\, 1.0$ and $1.5$, respectively. The time instants marked by different symbols in panel (a) correspond to those in panels (b)-(e).
	}
	\label{fig:shrinking}
\end{figure*}

This result has the following physical interpretation: the leading-order contribution to the MSD at short times is fully determined by the prescribed stochastic trajectory, i.e.,\ by the active driving protocol of the Ornstein-Uhlenbeck process \eqref{eq:active_eom}. In contrast, in the classical case, the scaling of the MSD is given by a Green–Kubo relation that involves velocity autocorrelations (see the SM):
\begin{equation}
    \text{MSD} = \int_0^t \int_0^t dt' dt'' \overline{\dot{x}(t')\dot{x}(t'')}.
\end{equation}

We note that, in the quantum case, the trapping potential exponent $q$ in Eq.~\eqref{eq:potential} enters the prefactor $A_q$ of Eq.~\eqref{eq:classical_average}, which is entirely quantum in origin. This prefactor determines whether short-time localization is stronger (negative 
$A_q$) or weaker (positive $A_q$). We remark that the emergence of a negative MSD is possible by virtue of its definition \eqref{eq:msd-define}, and its initial ground state condition. A negative MSD indicates that the wavefunction initially becomes narrower than its ground-state configuration, as is demonstrated in Fig.~\ref{fig:shrinking}(a). The mechanism is illustrated through time-sequential snapshots in Fig.~\ref{fig:shrinking}(b)–(e):\ on one side of the maximum of the wave function (or, more precisely, of $\langle \psi_t|\psi_t\rangle$), the moving potential localizes the wave function, essentially by pushing it, whereas on the other side, delocalization is negligible due to the low wavefunction amplitude on that side. Such negative MSD cannot occur for a classical stochastic particle with a sharp initial position. Even when the classical system is initiated with a non-zero temperature equilibrium configuration which has a non-zero initial MSD, the MSD increases with a $t^2$-scaling (see the SM, Sec.~\ref{sec:initial}). Thus, the occurrence of a negative MSD at short times is a pure quantum effect.

For a harmonic potential ($q=2$) the prefactor $A_q$ vanishes ($A_q = 0$), with the leading correction scaling as $t^6$ arising from second-order perturbation theory (see the SM, Sec.~\ref{subsection:harmonic}). For $q \ne 2$ (and thus $A_q \ne 0$), the leading-order contribution to the MSD \eqref{eq:classical_average} is $t^4$ (see the SM, Sec.~\ref{subsection:time_dependence}). In contrast, for the classical case, the potential exponent $q$ directly enters the dynamical exponent, resulting in a $t^{2(q+1)}$-scaling (see the SM, Sec.~\ref{sec:EOM}).

\begin{figure}[hb!]	
\includegraphics[width=0.95\columnwidth]{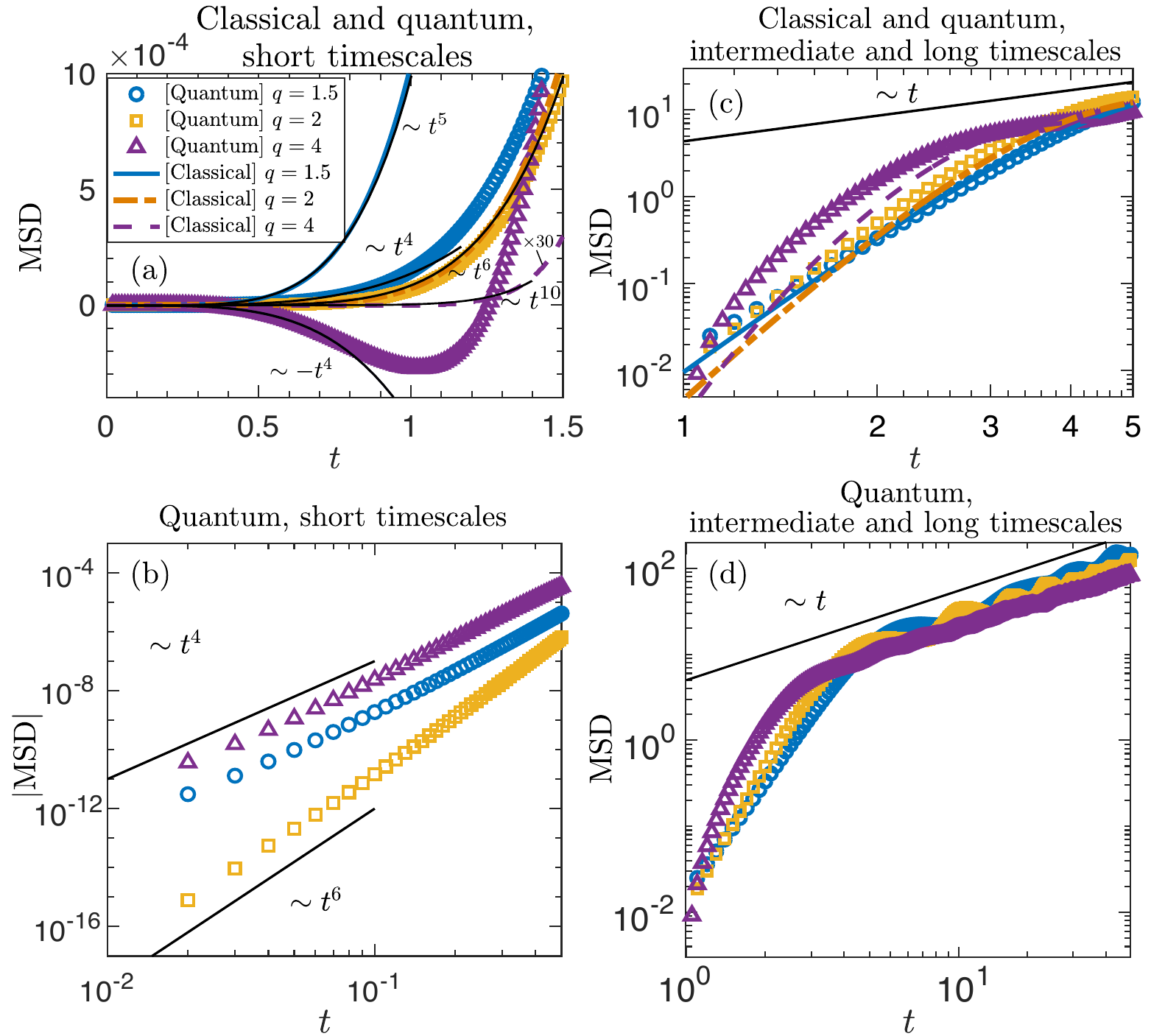}			
	\caption{{[\bf MSD of a non-dissipative quantum-active particle]}. (a) MSD at short timescales for $D=1$ and $\tau=10$ reveals quantum-active scaling being in general different from classical. (b) Log-log plot of the absolute values of MSD indicates $\sim t^4$ scaling for $q \neq 2$ and $\sim t^6$ for $q=2$.
 (c,~d) Crossover to the diffusive regime at long timescales for $D=1$ and $\tau=1$. (a,~c) Comparison of quantum and classical active MSD. In all panels, an average is taken over $500$ stochastic trajectories for each $q$ considered.}
	\label{fig:msd_non_dissipative}
\end{figure}

To test our analytical predictions, we now examine numerically in Fig.~\ref{fig:msd_non_dissipative}, how the shape of the trapping potential $V(x,t)$ influences the MSD for $q=1.5$, $q=2$, and $q=4$. In Fig.~\ref{fig:msd_non_dissipative}(a,~b), we see that for $q \ne 2$, the short-time quantum-active scaling of the MSD is indeed $\sim t^4$. For $q=4$, the MSD is initially negative, indicating that the quantum-active particle is transiently more localized than its ground state. Conversely, the MSD for $q=1.5$ scales positively as $\sim t^4$, rather than $\sim t^6$ as for the harmonic potential. At longer timescales ($t \gg \tau$), as expected, the MSD is fully determined by the trap motion, and the non-dissipative quantum particle enters the diffusive regime regardless of $q$ (see Fig.~\ref{fig:msd_non_dissipative}(c,~d)). Quantum-active effects become apparent when comparing the MSD to its classical counterpart, as the behavior differs at both shorter and longer timescales (Fig.~\ref{fig:msd_non_dissipative}(a,~c). Specifically, Fig.~\ref{fig:msd_non_dissipative}(a) highlights the characteristic $t^{2(q+1)}$-scaling of the classical particle. Thus, the $t$-dependence of the MSD between classical and active quantum particles is in general qualitatively different.

\section{Conclusion and discussions}

In this work, we introduced a generic approach for exploring quantum-active matter in continuum space. This approach involves mimicking the role of propulsion in active particles through imposing a trapping potential that is moving along imposed trajectories governed by classical active stochastic dynamics. We show that this framework for a single quantum degree of freedom, captures essential dynamical features of classical self-propelled active particles in the presence of any dissipation. These features include, a mean squared displacement (MSD) that scales diffusively ($\sim t$) at short times, ballistically ($\sim t^2$) at intermediate times, and again diffusively ($\sim t$) at times much longer than the persistence time $\tau$ of the active dynamics. Notably, this approach also reveals quantum-active dynamical features with no classical equivalent and that are most pronounced in the absence of dissipation. These features manifest as anomalous dynamical exponents of the MSD at short timescales. Most notably, the quantum-active $t^6$-scaling shown in Fig.~\ref{fig:msd_dissipative} is clearly observed in the range up to $10^{-3}-10^{-2}$, corresponding to typical particle displacements of $0.03 - 0.1$ in the oscillator length units. In the ultracold atom setup, where the oscillator length is typically $\sqrt{\hbar/m\omega} \sim 1$ \si{\micro\meter}, this translates to a physical displacement of about $30-100$  \si{\nano\meter}. Recent advances in quantum gas microscopy have enabled spatial resolutions down to 45  \si{\nano\meter} \cite{Veyron2024}, placing this quantum-active $t^6$-regime within the potential experimental observation.																																																																																	  

We stress that this work focuses on establishing a conceptually new and generic approach to study quantum-active matter that is meant to inspire future simulations and experiments. This approach is generalizable, and can be combined with or extended to interacting forms of classical active matter. This may include alternate forms of microscopic activity including run-and-tumble or active Brownian dynamics \cite{SolonCates2015, CatesTailleur2013, FreyKroy2005}. While the generalization to the many-body regime is beyond the scope of the present work, we would like to remark that: (i) experimental setups are now available to individually control more than $10^3$ quantum particles \cite{Lahaye2020}, which could be adapted in the future to mimic quantum active matter at the many-body level. (ii) In principle, a random but temporally and spatially correlated potential (a specific type of colored spatiotemporal noise) can be utilized to mimic quantum active motion at the many-body level at low to moderate densities and to predict a viable path toward realizing, e.g.,\ flocking and motility-induced phase separation. On the other hand, the effect of quantum phenomena such as exchange-symmetry of indistinguishable quantum particles (e.g. fermions or bosons) \cite{HyrkasManninen2013, AbrahamBonitz2012, ZhengPoletti2015, MyersDeffner2020}, or collective behavior of exotic quantum many-body ground-states (e.g. Bose-Einstein condensates) \cite{GreinerBloch2002}, on the dynamics of an externally activated process could also be interesting avenues to explore within the current context presented. Importantly, our theoretical predictions can be tested using state of the art experimental techniques in ultra-cold atoms confined to optical traps where the trap minimum can be programmed on prescribed stochastic trajectories. As such, our framework of active quantum matter could provide a paradigm for uncovering phenomenology involving the non-trivial coupling of quantum fluctuations with non-equilibrium features of activity. 

 \begin{acknowledgments}
 We thank I. Bloch, A. Hemmerich, P. Schmelcher,  A. Widera, J. Melles and R. Wittmann for helpful discussions. YZ acknowledges support from the Alexander-von-Humboldt Stiftung.
 \end{acknowledgments}

\clearpage
\widetext
\begin{center}
	\large{\textbf{Supplemental Material for \\ Engineering active motion in quantum matter}} 
	\normalsize
	
	\vspace{3ex}
	
	Alexander P.\ Antonov,$^1$ Yuanjian Zheng,$^1$ Benno Liebchen,$^2$ and Hartmut L\"owen$^1$
	\small
	\vspace{1ex}
	
	$^1$\textit{Institut f{\"u}r Theoretische Physik II: Weiche Materie, \\
		Heinrich-Heine-Universit{\"a}t D{\"u}sseldorf, 
		Universit{\"a}tsstra{\ss}e 1, 
		D-40225 D{\"u}sseldorf, 
		Germany}
	
	\vspace{.1ex}
	
	$^2$\textit{Technische Universit\"at Darmstadt, Hochschulstra{\ss}e 8, D-64289 Darmstadt, Germany}
\end{center}

\setcounter{equation}{0}
\setcounter{section}{0}
\setcounter{figure}{0}
\setcounter{table}{0}
\setcounter{page}{1}
\makeatletter
\renewcommand{\theequation}{S\arabic{equation}}
\renewcommand{\thefigure}{S\arabic{figure}}
\renewcommand{\bibnumfmt}[1]{[S#1]}
\renewcommand{\citenumfont}[1]{S#1}
\renewcommand{\thesection}{\Roman{section}}

\begin{center}
	\begin{minipage}{0.77\columnwidth}
		\small
		\hspace{2ex}
		In this supplementary document, we provide details to derive the key results in the main text and provide additional discussions of the quantum-active framework. We first discuss the non-dissipative quantum-active case in Sec.~\ref{sec:perturbation_theory}. In Sec.~\ref{sec:Lindblad}, we address the dissipative quantum-active case and provide details of the numerical calculations. Finally, in Sec.~\ref{sec:EOM}, we give an overview of classical active dynamics and derive the scaling laws for the mean squared displacement presented in the main text.
	\end{minipage}
\end{center}

\section{Non-dissipative Quantum Active Dynamics} 
\label{sec:perturbation_theory}
																																																																																																																																																																																																																																																																																																																 
In this section, we treat short time dynamics of the non-dissipative quantum-active particle. First, we review the basics of time-dependent perturbation theory in Sec.~\ref{subsection:perturbation_intro} before discussing in Sec.~\ref{subsection:perturbation}, the instructive case of first-order calculations for an initial pure (ground) state of the harmonic Hamiltonian. In doing so, we also derive the expression that relates the quantum mean-square-displacement (MSD) to a stochastic average of a classical ensemble, corresponding to Eq.~\eqref{eq:classical_average} in the main text. In Sec.~\ref{subsection:time_dependence}, we show how explicit time-dependence can be obtained from this expression, and further demonstrate that the lowest possible leading order dependence in time of the MSD is $\sim t^4$. This result is generalized to initial mixed-states in Sec.~\ref{subsection:mixed_states} and is shown to hold generally. This however is not the quantum-active scaling of the MSD for the special case of harmonic confining potentials, that instead scales as $\sim t^6$ as we show in Sec.~\ref{subsection:harmonic}. Lastly, in Sec.~\ref{sec:geometry} we provide additional numerical results for the family of power-law trapping potentials.

\subsection{Introduction to time-dependent perturbation theory}
\label{subsection:perturbation_intro}
We begin by reviewing basic results of time-dependent perturbation theory, following the pedagogy of \cite{Griffiths2018_supp}. First, consider a time-dependent Hamiltonian $\hat{H}(t)$ and corresponding potential $\hat{V}(\hat{x},t)$. The instantaneous state $\ket{\psi_t}_I$ evolves from an initial state $\ket{\psi_{t=0}}_I$ under a unitary operator $\hat{U}_I(t,t=0)$ in the interaction picture. The interaction or Dirac picture is a framework in quantum mechanics where the dynamics of a system are split between the state vectors and operators. It is commonly used for problems involving a solvable free Hamiltonian and a time-dependent perturbation term. The equation of motion (i.e. akin to the Schr{\"o}dinger equation in the Schr{\"o}dinger picture) is equivalently written:
\begin{equation}
	i\hbar \partial_t \ket{\psi_t}_I = \hat{V}_I(t) \ket{\psi_t}_I
	\label{eq:interaction_eom}
\end{equation}
where $\hat{V}_I=e^{i\hat{H}_{t=0}t/\hbar} \hat{V} e^{-i
	\hat{H}_{t=0} t/\hbar}$. By expanding $\ket{\psi_t}_I=\sum_n c_n \ket{n}$ in the initial energy orthogonal eigen-basis $ \{ \ket{n}\}$, with $\langle n|m\rangle = \delta_{nm}$, \eqref{eq:interaction_eom} reduces for a general state to:
\begin{equation}
	i\hbar \dot{c}_m = \sum_n V_{mn}(t)e^{i\frac{E_m-E_n}{\hbar}t} c_n(t)
\end{equation}
where $V_{mn}(t)=\braket{m|V(t)|n}$ and $E_n$ are the energy eigenvalues in the initial basis. Now, since unitary evolution $\ket{\psi_t}_I=\hat{U}_I\ket{\psi_{t=0}}_I$ holds similarly in the interaction picture, and \eqref{eq:interaction_eom} applies to any initial $\ket{\psi_{t=0}}_I$, it follows that
\begin{equation}
	i\hbar \partial_t \hat{U}_I(t,t=0) = \hat{V}_I(t) \hat{U}_I(t,t=0)
	\label{eq:U_dynamics}
\end{equation}
Upon solving, by time-integrating \eqref{eq:U_dynamics}, we obtain a closed-form expression for the unitary operator:
\begin{equation}
	\hat{U}_I(t,t=0)=\mathbb{I}-\frac{i}{\hbar}\int_{t=0}^t dt' \hat{V}_I(t')\hat{U}_I(t',t=0)
	\label{eq:time_integral}
\end{equation}
Note that the identity -- $\mathbb{I}$, comes from applying the (boundary) condition $\hat{U}_I(t=0,t=0)=\mathbb{I}$. 
\\
\indent
This expression can then be expanded by recursively substituting $\hat{U}$ within the time integral by its definition \eqref{eq:time_integral} to obtain a time-ordered integral:
\begin{align}
	\hat{U}_I(t,t=0)=\sum^{\infty}_{n=0} \left(-\frac{i}{\hbar}\right)^n \int_{t=0}^t dt_1 \dots  \int_{t=0}^{t_{n-1}} dt_{n} \hat{V}_I(t_1) \dots \hat{V}_I(t_n)
\end{align}
Lastly, since $\hat{U}_I(t,t=0) \ket{\psi_{t=0}}= \sum_n \ket{n}\bra{n}\hat{U}_I(t,t=0)\ket{\psi_{t=0}} \equiv \sum_n c_n(t) \ket {n}$, we arrive at explicit expressions for the (first and second order) corrections according to time-dependent perturbation theory:
\begin{align}
	c_n^{(1)}(t) &= - \frac{i}{\hbar}\sum_m\int dt'  e^{i\frac{(E_n-E_m)}{\hbar}t'} \braket{n | \hat{\tilde{H}}(t') |m}
	\\
	c_n^{(2)}(t) &= - \frac{1}{\hbar^2}\sum_m \int dt' \int dt''  e^{i\frac{(E_n-E_m)}{\hbar}t' + i\frac{(E_m-E_j)}{\hbar}t''} \braket{n | \hat{\tilde{H}}(t') |m}\braket{m | \hat{\tilde{H}} (t'') |j}
\end{align}
where $\hat{\tilde{H}}(t)= \hat{H}(t)-\hat{H}_{t=0}$ represents the perturbing change to the Hamiltonian at time $t$ equivalent to $\hat{V}_I$ in the interaction picture. In the following subsections, we shall make use of these two expressions to understand the short timescale dynamics of active quantum particles. 
\subsection{First-order perturbation theory}
\label{subsection:perturbation}

We consider a single non-relativistic quantum mechanical particle of mass $m$ in $1$-D that is subjected to an external harmonic potential - $V(x,t)=\frac{1}{2} m \omega^2 (x-x_c(t))^2$ of constant trap frequency $\omega$. This potential is being displaced along a prescribed random path $x_c (t)$ which defines the position of its potential minimum. The Hamiltonian is given explicitly by 
\begin{equation}
	\hat{H}(t)=\frac{\hat{p}^2}{2 m} +\frac{1}{2} m \omega^2 \left(\hat{x}-x_c(t)\right)^2
	\label{eq:hamiltonian}
\end{equation}
where $\hat{p}=-i\hbar\partial_{x}$ is the momentum operator.
We write the instantaneous Hamiltonian $\hat{H}(t)=\hat{H}_{t=0}+\hat{\tilde{H}}(t)$ as a time-dependent perturbation to initial Hamiltonian $\hat{H}_{t=0}$ , where
\begin{equation}
	\hat{\tilde{H}}(t) = \frac{1}{2} m \omega^2 \left[x_c^2(t)-2 \hat{x} x_c(t) \right]
\end{equation}
The probability of occupying the $j$-th eigenstate in the un-perturbed basis is then given by $P_j(t)=\vert c_j(t) \vert^2$, where
\begin{align}
	c_j(t) &=-\frac{i}{\hbar} \int_0^t dt' e^{i\frac{(E_j - E_0)}{\hbar}t'} \langle j \vert \hat{\tilde{H}} \vert 0 \rangle
	\\ &= - \frac{ m \omega^2 i}{2\hbar}   \int_0^t dt' e^{i\frac{(E_j - E_0)}{\hbar}t'} \left[x_c^2 \langle j \vert 0 \rangle - 2x_c \langle j \vert \hat{x} \vert 0 \rangle \right]
	\label{eq:c_m}
\end{align}
and $E_k$ is the eigenvalue of the $k$-th unperturbed energy level. The position operator is defined as $\hat{x}=\sqrt{\hbar/2m\omega}(\hat{a}^\dagger + \hat{a})$, where $\hat{a}^\dagger$ and $\hat{a}$ are the creation and annihilation ladder operators respectively, defined as $\hat{a}^\dagger \vert k \rangle=\sqrt{k+1} \vert k+1 \rangle$ and $\hat{a} \vert k \rangle=\sqrt{k} \vert k-1 \rangle$. Hence, we see that the prefactor associated to $x_c^2(t)$ only has surviving contributions coming from a transition to the ground state (i.e. $\langle 0 \vert 0 \rangle$), while the term linear in $x_c(t)$ has contributions originating only from excitation to its first excited state $\langle 1 \vert 0 \rangle$.

Furthermore, to leading order in $t$, the contribution from $x_c(t)$ dominates $x^2_c(t)$ in \eqref{eq:c_m}. Hence, the perturbative treatment of the time evolution is simply characterized by the $x_c(t)$ term in the occupation probability of the first excited state. 
\begin{equation}
	c_1(t)=i \sqrt{\frac{m\omega^3}{2\hbar}} \int_0^t dt' x_c(t') e^{i\omega t'}	
	\label{eq:first_excited}
\end{equation}
such that the global spread of the wavefunction can be approximated by 
\begin{equation}
	\langle \psi_t |\hat{x}^2|\psi_t\rangle  \sim \langle \psi_{t=0} |\hat{x}^2|\psi_{t=0}\rangle + \langle 1|\hat{x}^2|1 \rangle \vert c_1(t)\vert^2
\end{equation} 
where $\langle 1|\hat{x}^2|1 \rangle$ is the expectation value of $x^2$ for the first excited state of the initial Hamiltonian. Therefore, to leading order in $t$, the quantum mean-square displacement is given by:
\begin{equation}
	\langle \psi_{t} |\hat{x}^2|\psi_{t}\rangle - \langle \psi_{t=0} |\hat{x}^2|\psi_{t=0}\rangle\sim \left[ \int^t_0 dt' x_c(t')\right]^2
	\label{eq:spreading_wavefunction}
\end{equation}
and the effect of activity on $\langle \psi_t |\hat{x}^2|\psi_t\rangle$ associated to a given classical ensemble of trajectories $\left\{ \Gamma(t) : x_c(t) \right\}$  is revealed by considering the stochastic average of \eqref{eq:spreading_wavefunction}:
\begin{equation}
	\overline{\langle \psi_t |\hat{x}^2|\psi_t\rangle }= \frac{1}{\mathcal{Z}}\int d\Gamma 
		\langle \psi_t |\hat{x}^2|\psi_t\rangle_{\Gamma}
\end{equation}
where $\mathcal{Z}(\left\{ \Gamma \right\})$ denotes a normalization corresponding numerically to the number of paths drawn from a uniformly weighted ensemble. It then follows, that Eq.~\eqref{eq:classical_average} in the main text is recovered
\begin{equation}
	\overline{\langle \psi_{t} |\hat{x}^2|\psi_{t}\rangle} - \langle \psi_{t=0} |\hat{x}^2|\psi_{t=0}\rangle = A_{q=2}\overline{\left[ \int^t_0 dt' x_c(t')\right]^2} = A_{q=2}\int_0^t \int_0^t dt' dt'' \overline{x_c(t')x_c(t'')}.
	\label{eq:classical_average_supp}
\end{equation}
We note that this result holds for any $q>0$ in the generalized external potential $V(x,t)$ given by the equation:
	\begin{equation}
		V(x, t) = V_0 |x - x_c(t)|^q, \ V_0 = \frac{1}{2}\hbar^{1-\frac{q}{2}}m^{\frac{q}{2}}\omega^{\frac{q}{2}+1}, \ q > 0,
		\label{eq:potential-q}
	\end{equation}
	as the leading term in Eq.~\eqref{eq:c_m} is always linear in $x_c$.\ The corresponding calculations in Eqs.~\eqref{eq:first_excited}-\eqref{eq:classical_average_supp} are thus analogous, differing only in the prefactor $A_q$.

Finally, we reiterate that the result in Eq.~\eqref{eq:classical_average_supp} relates a quantum mechanical expectation value to a classical position auto-correlation function, which fundamentally reveals the nature of the quantum-active coupling in the current framework.

\subsection{Explicit time-dependence of the perturbative regime}
\label{subsection:time_dependence}

In this section, we derive the explicit lowest leading order $\sim t^4$ time-dependence for the quantum MSD in the perturbative quantum-active regime. For the dynamics of $x_c$ set by Eq.~\eqref{eq:active_eom} in the main text, the velocity ($\dot{x}_c$) auto-correlation function is given by \cite{Bonilla2019_supp, comment}:
\begin{equation}
	\overline{\dot{x}_c(t')\dot{x}_c(t'') }= \frac{D}{\tau} e^{- \vert t'-t'' \vert /\tau}
	\label{eq:velocity_correlation}
\end{equation}
where $D$ is the effective diffusive constant induced by the active force and $\tau$ is the persistence time of the active dynamics. Next, we write explicitly, the instantaneous position of the potential minimum $x_c(t)$ as a time-integral of its velocity for a given deterministic trajectory:
\begin{equation}
	x_c(t) =\int^t_0  dt' \dot{x}_c(t')
	\label{eq:x_c_as_time_integral}
\end{equation}
which leads to an expression for the \emph{position} auto-correlation function that is expressed as time-integrals of the imposed \emph{velocity} correlations for a given ensemble of stochastic trajectories:
\begin{equation}
	\overline{x_c(t_1)x_c(t_2)} = \int_0^{t_1} dt_3 \int_0^{t_2} dt_4 \overline{\dot{x}_c(t_3) \dot{x}_c(t_4)}= \int_0^{t_1} dt_3 \int_0^{t_2} dt_4 \frac{D}{\tau} e^{- \vert t_3-t_4 \vert /\tau}
	\label{eq:position_correlation}
\end{equation}
where the second equals sign in \eqref{eq:position_correlation} comes from using \eqref{eq:velocity_correlation}. Finally, by substituting \eqref{eq:position_correlation} into
\eqref{eq:classical_average_supp}, the stochastic average MSD of an initial ground state of an active quantum harmonic oscillator simply becomes:
\begin{equation}
	\overline{\langle \psi_t |\hat{x}^2|\psi_t\rangle} - \langle \psi_{t=0} |\hat{x}^2|\psi_{t=0}\rangle \sim \int_0^t dt_1 \int_0^t dt_2 \int_0^{t_1} dt_3 \int^{t_2}_0 dt_4 \frac{D}{\tau}  e^{-\vert t_3-t_4\vert/\tau},
\end{equation}
which to leading order in $t$ can easily be seen to lead to $\mathcal{O}(t^4)$ scaling behavior.

\subsection{Quantum-active scaling of excited states} 
\label{subsection:mixed_states}

Here, we generalize our perturbative treatment in Sec.~\ref{subsection:perturbation} to an initial wavefunction that involve excited states of the harmonic oscillator. i.e. the superposition $|\psi\rangle_{t=0}=\sum_k a_k \vert k_{t=0}\rangle$, where $\vert k_{t=0} \rangle$ is the energy eigenstates of the Hamiltonian $\hat{H}(t)$ at time $t=0$. Note that $a_k= e^{-\beta \hbar \omega(k+\frac{1}{2})}/\mathcal{Z}(\beta)$ for a Gibbs state in equilibrium at inverse temperature $\beta$, where $\mathcal{Z}=\sum_{k'}e^{-\beta \hbar \omega (k'+\frac{1}{2}))}$ is the partition function. In this scenario, the instantaneous occupation $c_j(t)$ of a general eigenmode $\vert j \rangle$ under first order perturbative expansion:
\begin{equation}
	c_j(t) =-\frac{i}{\hbar} \sum_k \int_0^t dt' e^{i\frac{(E_j - E_k)}{\hbar}t'} \langle j \vert \hat{\tilde{H}} \vert k \rangle
	\label{eq:weights}
\end{equation}
becomes relevant to the perturbative dynamics where $\hat{\tilde{H}}(t) = \frac{1}{2} m \omega^2 \left[x_c^2(t)-2 \hat{x} x_c(t) \right]$. However, since the position operator can be written $\hat{x}=\sqrt{\hbar/2m\omega}(\hat{a}^\dagger + \hat{a})$, contributions leading from $x^2_c(t)$ to \eqref{eq:weights} is non-zero only for $j=k$. Similarly, only $j=k\pm 1$ contributes to prefactors of the term linear in $x_c(t)$.
Hence, the inner product representing transition probabilities between $\vert k_{t=0} \rangle $ and $\vert j(t) \rangle$ can be written as:
\begin{equation}
	\langle j \vert \hat{\tilde{H}} \vert k \rangle =\frac{1}{2} m\omega^2 \left\{  x_c^2(t) \delta_{j,k} 
	\right.\\
	\left.
	- \sqrt{\frac{2\hbar}{{m\omega}}} \left[ \sqrt{k+1} \delta_{j,k+1} + \sqrt{k} \delta_{j,k-1} \right] x_c(t) \right\} 
	\label{eq:transition_probability}
\end{equation}
where $\delta_{p,q}$ denotes the Kronecker-delta function. As with in the case for the ground state considered in the main text, we note that dynamics leading from \eqref{eq:transition_probability} is likewise dominated by the $x_c(t)$ contribution, such that \eqref{eq:weights} can be expressed as
\begin{equation}
	c_j(t) \sim i \sqrt{\frac{m\omega^3}{2\hbar}}\int_0^t dt' x_c(t') \left[\sqrt{j} a_{j-1} e^{i\omega t'}+\sqrt{j+1}a_{j+1}e^{-i \omega t'}\right]
\end{equation}
which to leading-order in $t$, further simplify to
\begin{equation}
	c_j(t) \sim i \sqrt{\frac{m\omega^3}{2\hbar}}\left[\sqrt{j} a_{j-1}+\sqrt{j+1}a_{j+1}\right] \left[\int_0^t dt' x_c(t') \right]
	\label{eq:c_j_supp}
\end{equation}
where $a_j \equiv 0$ for $j < 0$. Note that \eqref{eq:c_j_supp} reduces to the ground state scenario (i.e.,\ Eq.~\eqref{eq:classical_average} in the main text) for $j=1$, $a_0=1$ and $a_{k \neq 0}=0$. Futhermore, since $\langle \psi_t |\hat{x}^2|\psi_t\rangle = \sum_j P_j \langle \hat{x}^2_j \rangle$ where $P_j=\vert c_j(t) \vert^2$ is the occupation probability and $\langle \hat{x}^2_j \rangle=\langle j|\hat{x}^2| j\rangle =\frac{\hbar}{m\omega}(j+\frac{1}{2})$ represents the spread of the $j$-th eigenmode respectively, the leading order behavior in $t$ is given by:
\begin{equation}
	\langle \psi_t |\hat{x}^2|\psi_t\rangle = \langle \psi_{t=0} |\hat{x}^2|\psi_{t=0}\rangle  + \frac{1}{2}\omega^2 \left[ \sum_j \left(j+\frac{1}{2}\right)\left( \sqrt{j} a_{j-1}+\sqrt{j+1}a_{j+1}\right)^2\right] \left[ \int_0^t dt' x_c(t')\right]^2
\end{equation}

Hence, we see that the ensemble average of the MSD for the general Gibbs state is still respecting the relation in Sec.~\ref{subsection:perturbation} for an initial ground state and Eq.~\eqref{eq:classical_average} in the main text.
In other words, the lowest leading-order quantum-active scaling $\sim t^4$  of the perturbative regime derived for an initial ground state, also holds for all Gibb's states of the harmonic oscillator, so long as the auto-correlation of the active velocity is prescribed by the active Ornstein-Uhlenbeck process.

\subsection{$t^6$-dependence for the harmonic oscillator}
\label{subsection:harmonic}

In this section, we elucidate the origins of the $\sim t^6$ leading order behavior of the MSD observed in Fig.~\ref{fig:msd_dissipative} and Fig.~\ref{fig:msd_non_dissipative} of the main text that is peculiar only to the harmonic oscillator. We begin by noting that the leading order contributions to the ground state occupation -- $c_0(t)$ comes from the complex phase $e^{-i\omega t/2}$ acquired due to the stationary dynamics such that:
\begin{equation}
	c_0(t) \sim  \left[ 1-\frac{1}{8}t^2 +\mathcal{O}(t^4) \right] + i\left[ - \frac{1}{2}t + \mathcal{O}(t^3) \right] 
\end{equation}
while the first-order perturbation to the first excited state -- $c^{(1)}_1(t)$ from \eqref{eq:first_excited} in Sec.~\ref{subsection:perturbation} can be written:
\begin{equation}
	c^{(1)}_1(t) \sim  \left[ -\frac{1}{3\sqrt{2}}t^3 +\mathcal{O}(t^5) \right] + i\left[ - \frac{1}{2\sqrt{2}}t^2 + \mathcal{O}(t^{4}) \right] 
\end{equation}
Now, for a pure state -- $ \ket{\psi}=\sum_n \ket{n}$, $\langle \hat{x}^2\rangle$ can be written as contributions from square ($ c_n^2$) and cross  $ (c^{\star}_m c_n )$ terms:
\begin{equation}
	\langle \psi|\hat{x}^2|\psi \rangle = \sum_{mn} (c^\star_m c_n + c_m c^\star_n)\braket{m | \hat{x}^2 | n} =\sum_n c_n^2 \langle \hat{x}_n^2 \rangle +\sum_{m \neq n} c^{\star}_m c_n \braket{m |\hat{x}^2 | n},
\end{equation}
where $c^*$ is a complex conjugate of $c$.\\
Hence, the square-terms contribution from first-order perturbation can be written:
\begin{equation}
	c_0^2 \langle \hat{x}_0^2 \rangle + c_1^2 \langle \hat{x}_1^2 \rangle = \left[ 1- \frac{1}{8}t^4+\mathcal{O}(t^8) \right] \langle \hat{x}_0^2 \rangle + \left[ 1 + \frac{1}{8}t^4+\mathcal{O}(t^6) \right] \cdot 3\langle \hat{x}_0^2 \rangle
	\label{eq:square_term_contribution}
\end{equation}
while the cross-terms are zero for first-order perturbations since $\braket{m| \hat{x}^2|m \pm 1}=0$. This implies that $1^{st}$ order perturbation theory alone predicts a non-vanishing $t^4$ leading order behavior for the MSD as also previously argued in Sec.~\ref{subsection:time_dependence}. Hence to understand the emergence of $\textrm{MSD} \sim t^6$ in the main text, we are compelled to consider $2^{nd}$-order perturbations. 

In $2^{nd}$-order time-dependent perturbation theory, the occupation of some higher excited state $\vert n \rangle$ can be written:
\begin{equation}
	c_n^{(2)}(t)= - \frac{1}{\hbar^2}\sum_m \int dt' \int dt''  e^{i\frac{(E_n-E_m)}{\hbar}t' + i\frac{(E_m-E_j)}{\hbar}t''} \braket{n | \hat{\tilde{H}}(t') |m}\braket{m | \hat{\tilde{H}} (t'') |j}
	\label{eq:2nd_order}
\end{equation}
In particular for the $2^{nd}$ excited state ($n=2$), we obtain
\begin{equation}
	c_2^{(2)}= \left[ -\frac{\sqrt{2}}{16} t^4+\mathcal{O}(t^6)\right] + i \mathcal{O}(t^5)
\end{equation}
which leads to cross-term contributions:
\begin{equation}
(c_0^\star c_2 + c_0 c_2^\star)\braket{0 | \hat{x}^2 | 2} =  2\sqrt{2} \left( \operatorname{Re}(c_2)\operatorname{Re}(c_0) + \operatorname{Im}(c_0)\operatorname{Im}(c_2)\right) \cdot\langle \hat{x}^2_0 \rangle = \left[- \frac{1}{4}t^4 +\mathcal{O}(t^6)\right]\langle \hat{x}_0^2\rangle.
\label{eq:cross_term_contribution}
\end{equation}
Combining \eqref{eq:square_term_contribution} and \eqref{eq:cross_term_contribution}, we see that the sum of $\mathcal{O}(t^4)$ contributions is zero, and hence the leading order behavior of the MSD for the harmonic oscillator is $\mathcal{O}(t^6)$.

\subsection{Numerical results}
\label{sec:geometry}

Here, we provide additional numerical results for the MSD of non-dissipative quantum-active particle at short time for various $q$ in the trapping potential $V(x,t)$ set by Eq.~\eqref{eq:potential-q}. As in the results presented in the main text, the numerical calculations are performed in real space using finite difference discrete time steps of $dt=10^{-1} - 10^{-2}$ and spatial resolution of the quantum mechanical $x$-spectrum is at least $dx=0.2$, while the MSD is averaged over $5 \times 10^2$ stochastic trajectories for each value of $q$ considered (see Sec.~\ref{sec:numerics} for the method).

\begin{figure}[htp!]	
	\includegraphics[width=0.8\columnwidth]{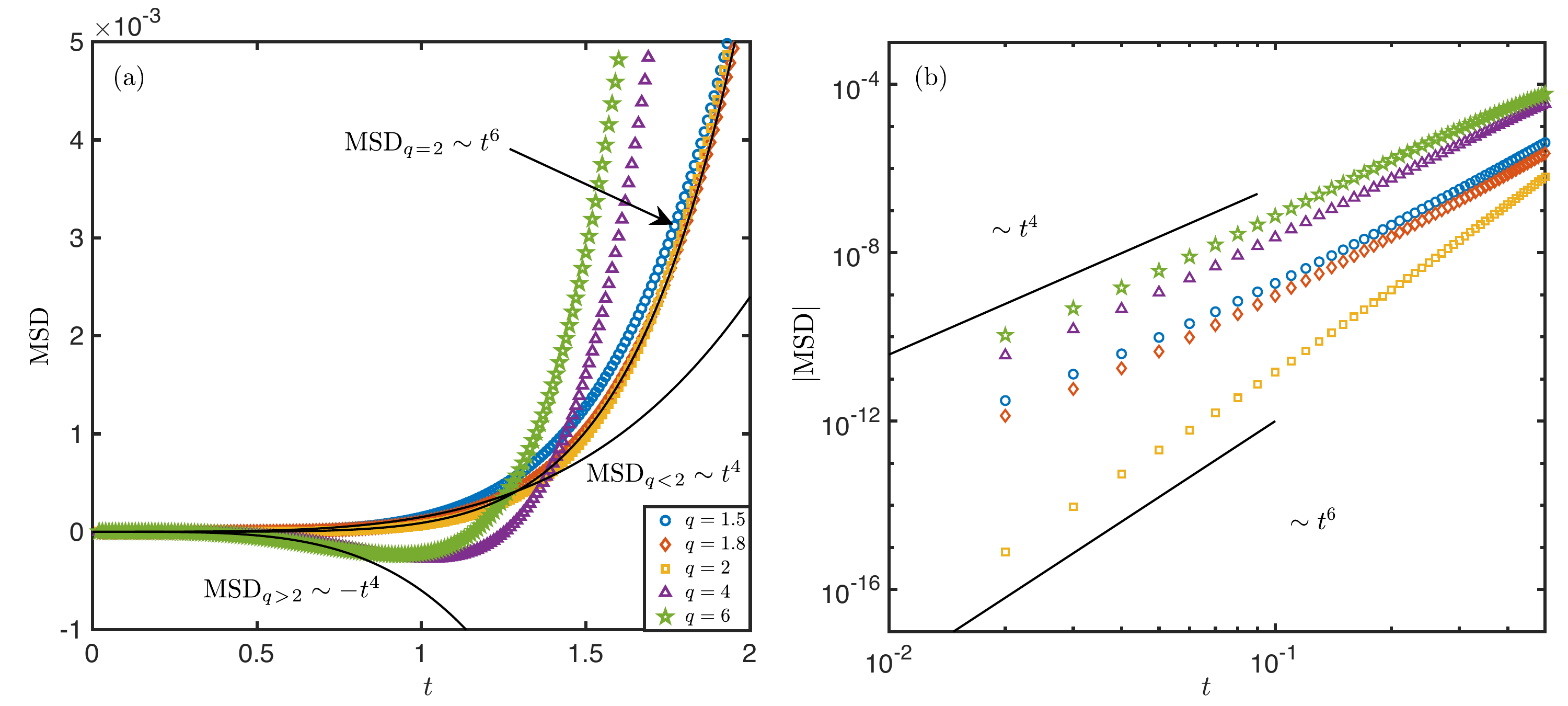}		
	\caption{{[\bf Trap Geometry dependence of the quantum MSD]}. (a) MSD and (b) $\vert\rm MSD\vert$ at short times for $D=1$, $\tau=10$, $dt=10^{-2}$ for $q=1.5,1.8,2,4,6$. }
	\label{fig:q}
\end{figure}
We see in Fig.~\ref{fig:q} that the $ \vert \textrm{MSD} \vert $ scales as $t^4$ for $q \neq 2$ as the perturbation theory Sec.~\ref{sec:perturbation_theory} predicts. Moreover the transient localization effect characterized by a negatively signed MSD for $q<2$ is also further verified, suggesting that these novel quantum-active features are intricately dependent on the exponent of the trapping potential.

\section{Dissipative Quantum Active Dynamics}
\label{sec:Lindblad}
In this section, we demonstrate the short-time diffusive behavior ($\sim t$) of the dissipative quantum-active particle. This behavior is analogous to thermal diffusion in the classical case and is notably absent in the non-dissipative scenario. First, in Sec.~\ref{sec:Lindblad_intro} we make a brief introduction to the Lindblad master equation that models dissipation of Markovian open quantum systems. In particular, we provide intuitive understanding to the different terms that appear in the equation, but defer its full derivation and details to an extensive existing literature \cite{RivasHuelga2011_supp,GardinerZoller2000_supp,BreuerPetruccione2002_supp,Manzano2020_supp}. In Sec.~\ref{sec:dissipation}, we demonstrate that the diffusive behavior observed at very short times originates from the dissipative term and is analogous to thermal diffusion in classical system.
\\
\subsection{Lindblad operators and open quantum systems}
\label{sec:Lindblad_intro}
\indent
The Lindblad master equation generally describes the time-evolution of a quantum system {\it weakly} coupled to a heat bath. It represents Markovian dissipation that does not require explicit consideration for the bath-system interactions, and is capable of generating non-unitary dynamics while still ensuring that the trace of the density matrix $\hat{\rho}(t)$ is preserved.
In general, the effective dynamics  can be written as an additional dissipative component $\mathcal{D}(\hat{\rho},t)$ to purely Hamiltonian dynamics of a closed system given by the von Neumann equation: 
\begin{equation}
	\dot{\hat{\rho}}=-\frac{i}{\hbar}\left[ \hat{H}(t),\hat{\rho}\right]+\hat{\mathcal{D}}(\hat{\rho},t)
	\label{eq:time_evolution_supp}
\end{equation}
Here, for simplicity we consider the dissipator composed of the Bosonic creation and annihilation operators $\hat{a}^\dagger$ and $\hat{a}$, which act respectively to either add or remove excitations to $\hat{\rho}(t)$ at rates controlled by $\nu_+$ and $\nu_-$ respectively. 
\begin{equation}
	\hat{\mathcal{D}}(\hat{\rho},t)=\nu_{+} \left(  \hat{a}^\dagger \hat{\rho} \hat{a} - \frac{1}{2}\left\{\hat{a}\hat{a}^\dagger ,\hat{\rho} \right\} \right) + \nu_{-} \left(  \hat{a} \hat{\rho} \hat{a}^\dagger - \frac{1}{2} \left\{\hat{a}^\dagger \hat{a} ,\hat{\rho} \right\} \right) \equiv \hat{\mathcal{D}}_+ +  \hat{\mathcal{D}}_-
	\label{eq:dissipator_supp}
\end{equation}
It should be noted that a for a moving external potential the creation and annihilation operators should be considered relative to the time-dependent position $x_c(t)$ of the potential minimum. We shall consider such a dynamical adjustment of the Lindblad damping in future work. This adjustment towards the co-moving frame will not alter the dynamical exponents reported in the present study.

To gain an intuition for the roles of these terms, we consider individually the effect of $\hat{\mathcal{D}}_+$ and $\hat{\mathcal{D}}_-$ for a general eigenstate $\ket{n}$ (i.e. $\hat{\rho}=\ket{n}\bra{n}$)
\begin{equation}
	\hat{\mathcal{D}}_+(\hat{\rho})  \sim  \hat{a}^\dagger \ket{n}\bra{n}\hat{a} - \frac{1}{2}\left(\hat{a}\hat{a}^\dagger\ket{n}\bra{n} + \ket{n}\bra{n}\hat{a}\hat{a}^\dagger \right) = (n+1)\ket{n+1}\bra{n+1} - (n+1)\ket{n}\bra{n} 
	\label{eq:dissipator_positive}
\end{equation}
\begin{equation}
	\hat{\mathcal{D}}_-(\hat{\rho})  \sim  \hat{a}\ket{n}\bra{n}\hat{a}^\dagger  - \frac{1}{2}\left(\hat{a}^\dagger\hat{a}\ket{n}\bra{n} + \ket{n}\bra{n}\hat{a}^\dagger \hat{a}\right) = n \ket{n-1}\bra{n-1} - n \ket{n}\bra{n}  
	\label{eq:dissipator_negative}
\end{equation}
Hence, we see that the effect of $\hat{\mathcal{D}}_+$ is essentially to excite the system by transferring occupation from $\ket{n}$ to $\ket{n+1}$, while conversely, $\hat{\mathcal{D}}_-$ destroys excitation by transferring occupation from  $\ket{n}$ to $\ket{n-1}$.
\\
\indent
Conceivably, for a given $\nu_+ < \nu_-$, the system reaches at long times to a steady state, characterized by a time-independent mean occupation $\langle \hat{n} \rangle$. In other words, $\partial_t\langle \hat{n}\rangle=0$, where $\langle n\rangle =\Tr\{\hat{a}^\dagger\hat{a} \hat{\rho}\} $ is the expectation of the number operator $\hat{n}=\hat{a}^\dagger\hat{a}$. For a time-independent Hamiltonian, this steady-state condition reduces the expectation value of \eqref{eq:dissipator_supp} to the relation
\begin{equation}
	\nu_- \langle \hat{n} \rangle=\nu_+(1+\langle \hat{n} \rangle)
\end{equation}
By further assuming that the heat bath is in thermal equilibrium at inverse temperature $\beta=1/k_B T$ such that the excitations are black-body (i.e. $\langle \hat{n} \rangle = (e^{\beta\hbar\omega}- 1)^{-1}$), we arrive at the expression that connects the bath temperature to the dissipation rates stated in the main text.
\begin{equation}
	\frac{\nu_+}{\nu_-}=e^{-\beta \hbar \omega }.
\end{equation}

\subsection{Short time diffusive dynamics}
\label{sec:dissipation}
In this section, we show that the initial diffusive regime discussed in the main text is always recovered in the presence of weak coupling to the environment when modeled as a Markovian Lindblad dissipator \eqref{eq:dissipator_supp}.
We consider $\langle \hat{x}^2 \rangle$ under the dynamics of $\hat{\mathcal{D}}$ for an infinitesimal $dt$ starting from a pure state $\hat{\rho}_0= \ket{n}\bra{n}$:
\begin{align}
	\frac{d\langle \hat{x}^2 (dt) \rangle}{dt} =\nu_{+} \left( \sum_k \braket{k \vert \hat{x}^2  \hat{a}^\dagger \hat{\rho}_0 \hat{a} \vert k}- \frac{1}{2}\sum_k \braket{k \vert \hat{x}^2 \hat{a}\hat{a}^\dagger \hat{\rho}_0 - \hat{x}^2 \hat{\rho}_0 \hat{a}\hat{a}^\dagger \vert k} \right) + \nu_{-} \left( \sum_k \braket{k \vert \hat{x}^2  \hat{a}\hat{\rho}_0 \hat{a}^\dagger  \vert k}- \frac{1}{2}\sum_k \braket{k \vert \hat{x}^2 \hat{a}^\dagger \hat{a}\hat{\rho}_0 - \hat{x}^2 \hat{\rho}_0 \hat{a}^\dagger 
		\hat{a}\vert k} \right)
	\label{eq:dissipation_msd}
\end{align}
The MSD contribution from the gain rate -- $\langle \hat{x}^2_{gain}\rangle$ reduces to 
\begin{equation}
	\frac{d\langle \hat{x}^2_{gain} \rangle}{dt} = \frac{\nu_+\hbar}{2 m \omega}\left[ (n+1)^2+ \frac{n+1}{2} \right]
	\label{eq:gain_rate}
\end{equation}
and similarly for  $\langle \hat{x}^2_{loss}\rangle$ associated to the loss rate,
\begin{equation}
	\frac{d\langle \hat{x}^2_{loss} \rangle}{dt} = \frac{\nu_-\hbar}{2 m \omega}n^2 
	\label{eq:loss_rate}
\end{equation}
Hence, we see that both contributions from the gain and loss terms for a pure state $\ket{n}$ to $\langle \hat{x}^2 \rangle$ are linear in $t$. For an initial ground state (i.e. $n=0$), \eqref{eq:loss_rate} is vanishing, such that only non-trivial contribution to the MSD comes from the gain rate \eqref{eq:gain_rate}, which upon integration yields
\begin{equation}
	\langle \hat{x}^2 \rangle = \frac{3 \nu_+\hbar}{4 m \omega} t
\end{equation}
More generally, we see that the same argument holds for any initial pure or mixed state at very short timescales, such that the MSD is always diffusive. For the quantum analogue of zero temperature ($\nu_+=0$), this diffusive behavior is absent, as demonstrated in Fig.~\ref{fig:msd_dissipative} of the main text.

\subsection{Numerical Details}
\label{sec:numerics}
\begin{figure}[htp!]	
	\includegraphics[width=0.5\columnwidth]{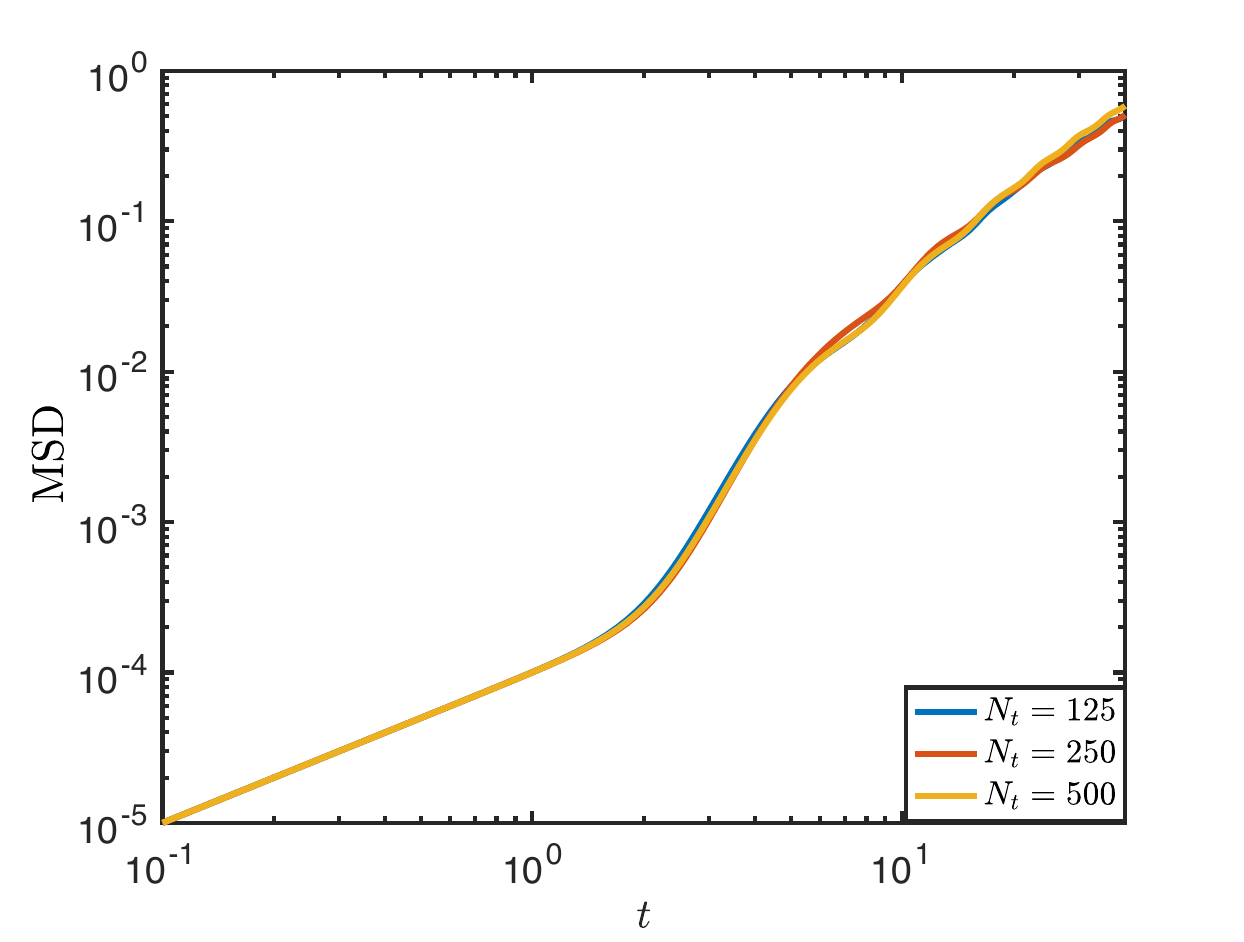}		
	\caption{{[\bf Convergence of the Dynamics]}. MSD averaged over various number of trajectories $N_t$. $D=0.01,\tau=10,q=2,\nu_-=10^{-2},\nu_+=10^{-4}$}
	\label{fig:convergence}
\end{figure}

In this section, we briefly discuss the numerical approach to solving the Lindblad equation \eqref{eq:time_evolution_supp} for a moving harmonic potential. First we generate an ensemble of trajectories -- $ \{ x_c(t)\}$ driven by simulating classical AOUPs for a given diffusion constant -- $D$ and persistence time -- $\tau$ that characterizes the degree of activity in the quantum particle. The velocity auto-correlation 	\eqref{eq:velocity_correlation} of the trajectories is ensured by simulating the corresponding microscopic equation of motion (Eq.~\eqref{eq:active_eom} in the main text):
\begin{equation}
	\tau\ddot{x}_c = -\dot{x}_c + \sqrt{2D}\eta_t,
	\label{eq:AOUP_0}
\end{equation}
where $\eta(t)$ is a zero-mean delta-correlated Gaussian white noise. This is solved using a first-order finite difference method with a time step $\delta t$ typically smaller than the physical time-step $\Delta t \sim 10^{-2}$ which is the discretized increment of time in the classical equation \eqref{eq:AOUP_0}. In practice, $\Delta t / \delta t= 5$.
\\
\indent
Now, for a given trajectory, $x_c(t)$ dictates the position of the minimum of a confining potential $V(x,t)$ \eqref{eq:potential-q} after each $\delta t$, which we use to simulate the time evolution of $\hat{\rho}$ under Lindblad dynamics using a predictor-corrector scheme \cite{numerical_supp}.
After each $\delta t$ a prediction for $\hat{\rho}(t+\delta t)$ is first calculated.
\begin{equation}
	\hat{\rho}_{pred}(t+\delta t)=\hat{\rho}(t)-\frac{i}{\hbar}[\hat{H}(x_c(t)),\hat{\rho}(t)] \delta t + \frac{\nu_-}{2}\left(2\hat{a}\hat{\rho}(t)\hat{a}^\dagger - \hat{a}^\dagger \hat{a}\hat{\rho}(t)  - \hat{\rho}(t)\hat{a}^\dagger \hat{a} \right)\delta t
	+ \frac{\nu_+}{2}\left(2\hat{a}^\dagger \hat{\rho}(t)\hat{a}- \hat{a}\hat{a}^\dagger \hat{\rho}(t)  - \hat{\rho}(t)\hat{a}\hat{a}^\dagger\right)  \delta t 
\end{equation}
Subsequently, this prediction $\hat{\rho}_{pred}$ is combined with $\hat{\rho}(t)$ in the corrector step:
\begin{equation}
	\hat{\rho}_m=\frac{1}{2}(\hat{\rho}(t)+\hat{\rho}_{pred}(t+\delta t))
\end{equation}
\begin{equation}
	\hat{\rho}(t+\delta t)=\hat{\rho}(t)-\frac{i}{\hbar}[\hat{H}(x_c(t)),\hat{\rho}_m(t)] \delta t + \frac{\nu_-}{2}\left(2\hat{a}\hat{\rho}_m(t)\hat{a}^\dagger - \hat{a}^\dagger \hat{a}\hat{\rho}_m(t)  - \hat{\rho}_m(t)\hat{a}^\dagger \hat{a} \right) \delta t
	+ \frac{\nu_+}{2}\left(2\hat{a}^\dagger \hat{\rho}_m(t)\hat{a}- \hat{a}\hat{a}^\dagger \hat{\rho}_m(t)  - \hat{\rho}_m(t)\hat{a}\hat{a}^\dagger \right)\delta t
\end{equation}
To obtain the mean-square displacement (MSD) at $t$ along a single given trajectory $x_c(t)$, we calculate:
\begin{equation}
	\textrm{MSD}[x_c] = \textrm{Tr}\left({\hat{\rho}(t)\hat{x}^2}\right) - \textrm{Tr}\left({\hat{\rho}_{t = 0}\hat{x}^2}\right).
\end{equation}
which is then subsequently averaged over the stochastic trajectories for a given $D$ and $\tau$. In Fig.~\ref{fig:convergence}, we show the behavior of the average MSD for various ensemble sizes, and note that they converge at relatively few number of trajectories (compared to classical active particle simulations). This could be a consequence of the identical initial conditions (i.e. quantum (ground) state), the lower dimensionality (1D) of the space, and by virtue of the dynamics being a single particle. 

\newpage
\section{Classical active dynamics}
	\label{sec:EOM}
	In this section, we provide a brief introduction to active motion. In particular, we consider the active Ornstein-Uhlenbeck process \eqref{sec:AOUP}, for which we provide analytical calculation of the MSD scaling \eqref{sec:AOUP-short}. It is followed by an analytical calculation of the MSD scaling for a passive particle driven with a moving trapping potential \eqref{sec:mimicking-active} to mimic classical active motion. In Table~\ref{table:eom}, we summarize classical equations of motion corresponding to self-propelled particle and particle mimicking self-propulsion. In Sec.~\ref{sec:initial}, we consider the MSD of initially distributed particles.
	\begin{table}[h!]
		\begin{center}
			\begin{tabular}[c]{||c || c || c || c || }
				\hline
				& \specialcell{\vspace{-2ex} \\ Classical Self-Propelled Particle\vspace{0.5ex} \\ (AOUP)} & \specialcell{\vspace{-2ex} \\Mimicking Classical\vspace{0.5ex} \\ Self-propulsion} \\
				\hline
				\specialcell{\vspace{-2ex} \\ Dissipative \\ (overdamped)} & \specialcell{\vspace{-2ex} \\ $\dot{x} = \dot{x}_c + \sqrt{2D_t}\zeta(t)$ \\ Main text, Fig.~\ref{fig:schematic_dissipative}, left column} & \specialcell{\vspace{-2ex} \\ $\gamma\dot{x} = -\nabla V(x,t) + \gamma\sqrt{2D_t}\zeta(t)$ \\ Main text, Fig.~\ref{fig:schematic_dissipative}, center column} \\
				\hline
				\specialcell{\vspace{-2ex} \\ Non-dissipative \\ (inertial)} & \specialcell{\vspace{-2ex} \\ $m\ddot{x} = \gamma_c\dot{x}_c$ \vspace{0.5ex}\\ Listed for completeness; Sec.~\ref{sec:EOM}\hspace{.2ex}B} & \specialcell{\vspace{-2ex} \\ $m\ddot{x} = -\nabla V(x,t)$ \\ Main text, Fig.~\ref{fig:schematic_non_dissipative}, right column} \\
				\hline
			\end{tabular}
		\end{center}
		\caption{Classical active equations of motion for self-propelled particle and particle mimicking self-propulsion. Here, $m$ is the particle mass, $\gamma$ is a damping coefficient, $D_t$ is the thermal diffusion constant, $V(x,t) = V_0|x-x_c(t)|^q$ is the driving potential \eqref{eq:potential-q}, $\zeta(t)$ is a zero-mean delta-correlated Gaussian white noise and $\gamma_c$ is the active force amplitude. For a more detailed technical review, see Refs.~\cite{Bonilla2019_supp, MartinvanWijland2021_supp, nguyen2021active_supp}.}
		\label{table:eom}
	\end{table}
	\subsection{Introduction to active motion}
	\label{sec:AOUP}
	The dynamics of active particles are commonly described by the stochastic equation:
	\begin{equation}
		m\ddot{x} + \gamma\dot{x} = f_a +\!\sqrt{2D_t}\zeta(t),
		\label{eq:active-particle}
	\end{equation}
	where $m$ is the particle mass, $\gamma$ is the damping coefficient, $D_t$ is the thermal diffusion coefficient, $\zeta(t)$ is a zero-mean delta-correlated Gaussian white noise and $f_a$ is the represents the active force distinguishing active dynamics from passive Brownian motion.
	The Active Ornstein-Uhlenbeck Particle (AOUP) model describes self-propelled particles where the active force $f_a$ evolves according to an Ornstein-Uhlenbeck process (Ornstein-Uhlenbeck) process \cite{Bonilla2019_supp, MartinvanWijland2021_supp, nguyen2021active_supp}: 
	\begin{equation}
		\tau\dot{f_a} = -f_a + \gamma_c\!\sqrt{2D}\eta(t),
		\label{eq:AOUP_force}
	\end{equation}
	where $\tau$ is the persistence time, $D$ is the diffusive coefficient induced by the active force, $\gamma_c$ is the amplitude of the active force $f_a = \gamma_c \dot{x}_c$ and $\eta(t)$ is a zero-mean delta-correlated Gaussian white noise. In terms of positional variable $x_c$, Eq.~\eqref{eq:AOUP_force} is identical to Eq.~\eqref{eq:active_eom} of the main text:
	\begin{equation}
		\tau\ddot{x}_c = -\dot{x}_c +\!\sqrt{2D}\eta(t).
		\label{eq:AOUP}
	\end{equation}
	The AOUP is widely used to study active matter because it effectively captures persistency of active motion while maintaining both mathematical and computational simplicity. \\
	\subsection{Short time AOUP dynamics}
	\label{sec:AOUP-short}
																																																								
	In the overdamped limit $m \to 0$, Equation~\eqref{eq:active-particle} with $\gamma = \gamma_c$ reduces to:
	\begin{equation}
		\dot{x} = \dot{x}_c +\!\sqrt{2D_t}\zeta(t).
	\end{equation}
	This limit corresponds to the left column in Fig.~\ref{fig:schematic_dissipative} of the main text. Here, we calculated the MSD with the Green-Kubo relation:
	\begin{equation}
		\textrm{MSD} = \int_0^t\int_0^t dt' dt'' \overline{\dot{x}(t')\dot{x}(t'')} =  2D_t \int_0^t\int_0^t dt' dt'' \overline{\zeta(t')\zeta(t'')} + \int_0^t\int_0^t dt' dt'' \overline{\dot{x}_c(t')\dot{x}_c(t'')}. 
	\end{equation}
	For the velocity-correlation function of $x_c$ given by Eq.~\eqref{eq:velocity_correlation}, we get
	\begin{equation}
		\textrm{MSD}= 2D_t \int_0^t\int_0^t dt' dt'' \overline{\zeta(t')\zeta(t'')} + \int_0^t\int_0^t dt' dt'' \overline{\dot{x}_c(t')\dot{x}_c(t'')} = 2 D_t t +  2D(t + \tau[e^{-t/\tau} - 1]).
		\label{eq:MSD-active}
	\end{equation}
	Equation~\eqref{eq:MSD-active} is given by the sum of two contributions: i) a thermal noise contribution, $2D_t t$, and ii) a contribution due to the active force, which is $\propto D$. For $t \ll \tau$, we obtain
	\begin{equation}
		\textrm{MSD} \approx 2 D_t t + \frac{D}{\tau}t^2.
	\end{equation}
	For short timescales the diffusive regime $\sim t$ is solely controlled by the thermal noise amplitude $D_t$. This situation is analogous to a dissipative quantum case discussed in Sec.~\ref{sec:dissipation}, where the diffusive regime at short timescales is governed by the gain rate $\nu_+$ associated with the equilibrium temperature of the thermal bath.\\
	A ballistic regime $\sim t^2$ due to the active force emerges at intermediate times, while for the long timescales $t \gg \tau$, the dynamics transitions back to a diffusive behavior:
	\begin{equation}
		\textrm{MSD} \approx 2(D_t + D)t.
	\end{equation}
	For the inertial (undamped) limit $\gamma = 0$, Equation \eqref{eq:active-particle} simplifies to
	\begin{equation}
		m\ddot{x} = \gamma_c \dot{x}_c.
		\label{eq:t-4}
	\end{equation}
	Similar to the quantum case discussed in Sec.~\ref{sec:Lindblad}, we now express $x_c(t)$ as a time integral of its velocity \eqref{eq:x_c_as_time_integral} in the Green-Kubo relation, and obtain an expression for the classical MSD:
	\begin{equation}
		\textrm{MSD} = \displaystyle \frac{\gamma_c^2}{m^2}\int_0^t dt_1 \int_0^{t_1} dt_2 \int_0^{t}dt_3 \int_0^{t_3} dt_4 \frac{D}{\tau}e^{-|t_2-t_4|/\tau} = \frac{D\gamma_c^2}{3 m^2}\left(2t^3 - 3t^2\tau + 6\tau^3 - 6e^{-t/\tau}\tau^2(t+\tau) \right). 
	\end{equation}
	It exhibits quartic scaling for short timescales $t \ll \tau$,
	\begin{equation}
		\textrm{MSD} \approx \frac{D\gamma_c^2}{4\tau  m^2}t^4,    
	\end{equation}
	while for long timescales $t \gg \tau$, it follows the cubic scaling: 
	\begin{equation}
		\textrm{MSD} \approx \frac{2 D\gamma_c^2}{3 m^2}t^3.
	\end{equation}
	\subsection{Mimicking active motion}
	\label{sec:mimicking-active}
	Let us now consider a passive classical particle that is set in motion via a moving trapping potential $V(x,t)$, Eq.~\eqref{eq:potential-q}. The resulting dynamics read:
	\begin{equation}
		m\ddot{x} + \gamma\dot{x} = -\nabla V(x,t) + \gamma\sqrt{2D_t}\zeta(t).
		\label{eq:driving}
	\end{equation}
	This type of driving is effective for approximating the properties of the active motion, since (i) the thermal diffusive regime $\sim t$ persists in the presence of the white noise term, and (ii) it recovers the MSD scalings for intermediate $\sim t^2$ and long $\sim t$ timescales. Thus, a passive particle in the moving trapping potential mimics the essential features of AOUP.\\
	However, the imposed potential introduces an additional scaling regime for short timescales.\\
	In the overdamped limit $m\to 0$ with $\gamma = \gamma_c$ and harmonic potential $q=2$, Equation~\eqref{eq:driving} reduces to
	\begin{equation}
		\gamma\dot{x} = -m_c \omega^2(x - x_c).
	\end{equation}
	Since $x_c$ dominates $x$ for $t \ll \tau$, at short timescales this equation is structurally identical to Eq.~\eqref{eq:t-4}, thus giving us $t^4$-scaling in between the diffusive regime at very short times and ballistic regime at intermediate times (center column of Fig.~\ref{fig:schematic_dissipative} in the main text).
	For the inertial limit $\gamma \to 0$ with $m = m_c$ and harmonic potential $q=2$, \\
	\begin{equation}
		\ddot{x} + \omega^2 x = \omega^2 x_c(t),
	\end{equation}
	which is a non-homogeneous second-order linear differential equation representing an undamped driven harmonic oscillator. For the particle initialized in the potential minimum $x(0) = x_c(0) = 0$ (corresponding to the ground state in the quantum case), the solution is:  
	\begin{equation}
		x(t) = \text{Re} \left[ \omega^2 \int_0^t dt_1 \int_0^{t_1} dt_2 \, x_c(t_2) \, e^{i\omega[(t_2 - t_1) + (t - t_1)]} \right].
		\label{eq:formal_solution}
	\end{equation} 
	We present $x_c(t)$ as a time integral of its velocity \eqref{eq:x_c_as_time_integral} in the solution \eqref{eq:formal_solution}, with \eqref{eq:velocity_correlation} as the active velocity correlation, and obtain the following expression for the classical MSD:
	\begin{equation}
		\langle x^2(t) \rangle \sim \omega^4 \int_0^t dt_1 \int_0^{t_1} dt_2 \int_0^{t_2} dt_3 \int_0^t dt_4 \int_0^{t_4} dt_5 \int_0^{t_5} dt_6 \frac{D}{\tau} e^{-|t_3 - t_6|}.
	\end{equation}   
	To the lowest order in $t$, this reduces to:  
	\begin{equation}
		\langle x^2(t) \rangle \sim \omega^4 \frac{D}{36\tau} t^6 + O(t^8),
	\end{equation}  
	corresponding to the short-time sextic scaling discussed in the main text (left column in Fig.~\ref{fig:schematic_non_dissipative} for $q = 2$).\\
	For an arbitrary $q$ in the external potential $V(x,t) = V_0 |x - x_c(t)|^q$, since $x_c$ dominates $x$ in the equation of motion for short times, we have
	\begin{equation}
		x(t) \sim \int_0^t dt_1 \int_0^{dt_1} dt_2 [x_c(t_2)]^{q-1},
	\end{equation}
To generalize Eq.~\eqref{eq:x_c_as_time_integral}, we use the Fa\`a di Bruno formula for the $(q-1)$-th derivative of $[x_c(t)]^{q-1}$,
\begin{equation}
	\frac{d^{q-1}}{dt^{q-1}} \left[ x_c(t)\right]^{q-1} 
	=
	\sum_{k=0}^{q-1}
	\frac{(q-1)!}{(q-1 - k)!}
	\, [x_c(t)]^{q-1 - k}
	\, B_{q-1,k}\left(\dot{x}_c(t), \ddot{x}_c(t), \dots, x_c^{(q - k)}(t)\right)
	\label{eq:faa}
\end{equation}
where $B_{q-1,k}$ are the Bell polynomials in the derivatives of $x_c(t)$ \cite{combinatorics_supp}. In Equation~\eqref{eq:faa}, term $B_{q-1,q-1} = (q-1)! [\dot{x}_c(t)]^{q-1}$ for $k = q-1$ dominates the contributions from higher-order derivatives at short times. This leads to a generalization of Eq.~\eqref{eq:x_c_as_time_integral}:
	\begin{equation}
		[x_c(t)]^{q-1} \sim \int_0^t dt_1 \int_0^{t_1} dt_2 \ldots \int_0^{t_{q-2}} dt_{q-1} [\dot{x}_c(t_q)]^{q-1},
	\end{equation}
	we obtain
	\begin{equation}
		\langle x^2(t)\rangle \sim \int_0^t dt_1 \int_0^{t_1} dt_2 \ldots \int_0^{t_{q}} dt_{q+1} \int_0^t dt_{q+2} \int_0^{t_{q+2}} dt_{q+3} \ldots \int_0^{t_{2q+1}} dt_{2q+2} \frac{D}{\tau} e^{-q|t_{q+1} - t_{2q+2}|} \sim t^{2(q+1)},
		\label{eq:q-scaling}
	\end{equation}
	which is the result for short timescales for $q > 0$ that is shown in the left column of Fig.~\ref{fig:schematic_non_dissipative} in the main text.
	\subsection{Short time mean squared displacement of initially distributed particles.}
	\label{sec:initial}
	For a more general scenario, where the initial position and velocity of the particle are non-zero, the equation of motion can be written as the integral equation
	\begin{equation}
		x(t) = v(0)t + x(0) - \int_0^t dt_1 \int_0^{t_1} \partial_x V(x(t_2), t_2) dt_2,
	\end{equation}
	where the integral term is dominated by the initial conditions $x(0), v(0)$, at short times. We thus obtain the following expression:
	\begin{equation}
		\langle x^2(t)\rangle - \langle x^2(0)\rangle \sim \langle v^2(0) \rangle t^2 + 2 \langle x(0) \rangle \langle v(0) \rangle t.
	\end{equation}
For an initial equilibrium Maxwell–Boltzmann distribution of velocities at finite temperature, and a Boltzmann distribution of positions corresponding to the potential given in Eq.~\eqref{eq:potential-q}, 
\begin{subequations}
	\begin{eqnarray}
		\langle x(0) \rangle & = & 0,\\
		\langle v(0) \rangle & = & 0,\\
		\langle v^2(0) \rangle & = & \frac{3 k_B T}{m},\\
		\langle \partial_x V(x(0)) x(0) \rangle & = & \frac{V_0q}{m} \frac{\int_{-\infty}^{\infty} e^{-\frac{V_0|x|^q}{k_B T}} |x|^q dx}{\int_{-\infty}^{\infty} e^{-\frac{V_0|x|^q}{k_B T}} dx} = \frac{k_B T}{m},
	\end{eqnarray}
\end{subequations}
and the classical $\textrm{MSD} = \langle x^2(t)\rangle - \langle x^2(0)\rangle$ is given by:
\begin{equation}
	\textrm{MSD} \sim \frac{2 k_{\rm B}T}{m}t^2,
\end{equation}
Therefore, a classical system with an initial equilibrium distribution at finite temperature always has a scaling that is different from the one obtained in the quantum case. In addition, more important, the negative MSD observed in the non-dissipative quantum case is not present in classical systems at finite temperatures.

\bibliographystyle{unsrt}

\begin{thebibliography}{89}


\bibitem{MarchettiSimha2013} M. Marchetti, J. Joanny, S. Ramaswamy, T. Liverpool, J. Prost, M. Rao, and R. A. Simha, {\it Hydrodynamics of soft active matter}, Rev. Mod. Phys. {\bf 85}, 1143 (2013).

\bibitem{Ramaswamy2010} S. Ramaswamy, {\it The mechanics and statics of active matter}, Annu. Rev. Condens. Matter Phys. {\bf 1}, 323–345 (2010).

\bibitem{Gompper2020} G. Gompper et al., {\it The 2020 Motile Active Matter Roadmap}, J. Phys. Condens. Matter {\bf 32}, 193001 (2020).

\bibitem{paxton2004catalytic}
W. F. Paxton, K. C. Kistler, C. C. Olmeda, A. Sen, S. K. St. Angelo, Y. Cao, T. E. Mallouk, P. E. Lammert, and V. H. Crespi,
\textit{Catalytic nanomotors: autonomous movement of striped nanorods},
J. Am. Chem. Soc. \textbf{126}, 13424 (2004).

\bibitem{BechingerVolpe2016} C. Bechinger, R. Di Leonardo, H. L{\"o}wen, C. Reichhardt, G. Volpe, and G. Volpe, { \it Active Particles in Complex and Crowded Environments}, Rev. Mod. Phys. {\bf 88}, 045006 (2016).


\bibitem{howse2007self}
J. R. Howse, R. A. Jones, A. J. Ryan, T. Gough, R. Vafabakhsh, and R. Golestanian,
\textit{Self-motile colloidal particles: from directed propulsion to random walk},
Phys. Rev. Lett. \textbf{99}, 048102 (2007).


\bibitem{DreyfusBibette2005} R. Dreyfus , J. Baudry, M. L. Roper, M. Fermigier, H. A. Stone, J. Bibette, {\it Microscopic artificial swimmers}, Nature (London) {\bf 437}, 862 (2005).

\bibitem{TiernoSagues2008} P. Tierno, R. Golestanian, I. Pagonabarraga, and F. Sagues, {\it Magnetically Actuated Colloidal Microswimmers}, J. Phys. Chem. B {\bf 112}, 16525 (2008).

\bibitem{DeseigneChate2010} J. Deseigne, O. Dauchot, and H. Chat{\'e}, {\it Collective motion of vibrated polar disks}, Phys. Rev. Lett. {\bf 105}, 098001 (2010).

\bibitem{LimJaeger2019} M. X. Lim, A. Souslov, V. Vitelli, and H. M. Jaeger, {\it Cluster formation by acoustic forces and active fluctuations in levitated granular matter}, Nat. Phys. {\bf 15}, 460 (2019).


\bibitem{schmidt2019light}
F. Schmidt, B. Liebchen, H. Löwen, and G. Volpe,
\textit{Light-controlled assembly of active colloidal molecules},
J. Chem. Phys. \textbf{150}, 9 (2019).

\bibitem{saha2020scalar}
S. Saha, J. Agudo-Canalejo, and R. Golestanian,
\textit{Scalar active mixtures: The nonreciprocal Cahn-Hilliard model},
Phys. Rev. X \textbf{10}, 041009 (2020).

\bibitem{ivlev2015statistical}
A. V. Ivlev, J. Bartnick, M. Heinen, C-R. Du, V. Nosenko, and H. Löwen,
\textit{Statistical mechanics where Newton’s third law is broken},
Phys. Rev. X \textbf{5}, 011035 (2015).


\bibitem{grauer2021active}
J. Grauer, F. Schmidt, J. Pineda, B. Midtvedt, H. Löwen, G. Volpe, and B. Liebchen, \textit{Active droploids}, Nat. Commun. \textbf{12}, 6005 (2021).

\bibitem{SurowkaBanerjee2023} P. Sur{\'o}wka, A. Souslov, F. J{\"u}licher, and D. Banerjee, {\it Odd Cosserat elasticity in active materials}, Phys. Rev. E {\bf 108}, 064609 (2023).

\bibitem{soto2014self}
R. Soto and R. Golestanian,
\textit{Self-assembly of catalytically active colloidal molecules: Tailoring activity through surface chemistry},
Phys. Rev. Lett. \textbf{112}, 068301 (2014).

\bibitem{niu2018dynamics}
R. Niu, A. Fischer, T. Palberg, and T. Speck,
\textit{Dynamics of binary active clusters driven by ion-exchange particles},
ACS Nano \textbf{12}, 10932 (2018).

\bibitem{toner1995long}
J. Toner and Y. Tu,
\textit{Long-range order in a two-dimensional dynamical XY model: how birds fly together},
Phys. Rev. Lett. \textbf{75}, 4326 (1995).


\bibitem{toner1998flocks}
J. Toner and Y. Tu,
\textit{Flocks, herds, and schools: A quantitative theory of flocking},
Phys. Rev. E \textbf{58}, 4828 (1998).

\bibitem{narayan2007long}
V. Narayan, S. Ramaswamy, and N. Menon,
\textit{Long-lived giant number fluctuations in a swarming granular nematic},
Science \textbf{317}, 105 (2007).
\bibitem{cates2015motility}
M. E. Cates and J. Tailleur,
\textit{Motility-induced phase separation},
Phys. Rev. Lett. \textbf{6}, 219 (2015).


\bibitem{theurkauff2012dynamic}
I. Theurkauff, C. Cottin-Bizonne, J. Palacci, C. Ybert, and L. Bocquet,
\textit{Dynamic clustering in active colloidal suspensions with chemical signaling},
Phys. Rev. Lett. \textbf{108}, 268303 (2012).

\bibitem{palacci2013living}
J. Palacci, S. Sacanna, A. P. Steinberg, D. J. Pine, and P. M. Chaikin,
\textit{Living crystals of light-activated colloidal surfers},
Phys. Rev. Lett. \textbf{339}, 936 (2013).

\bibitem{buttinoni2013dynamical}
I. Buttinoni, J. Bialké, F. Kümmel, H. Löwen, C. Bechinger, and T. Speck,
\textit{Dynamical clustering and phase separation in suspensions of self-propelled colloidal particles},
Phys. Rev. Lett. \textbf{110}, 238301 (2013).

\bibitem{ginot2018aggregation}
F. Ginot, I. Theurkauff, F. Detcheverry, C. Ybert, and C. Cottin-Bizonne,
\textit{Aggregation-fragmentation and individual dynamics of active clusters},
Phys. Rev. Lett. \textbf{9}, 696 (2018).


\bibitem{Whitesides2018} G. M. Whitesides, {\it Soft robotics}. Angew. Chem. Int. Ed. {\bf 57}, 4258–4273 (2018).

\bibitem{LeymanVolpe2018} M. Leyman, F. Ogemark, J. Wehr, and G. Volpe, {\it Tuning phototactic robots with sensorial delays}, Phys. Rev. E {\bf 98}, 052606 (2018).

\bibitem{VasarhelyiVicsek2018} G. V{\'a}s{\'a}rhelyi, C. Vir{\'a}gh, G. Somorjai, T. Nepusz, A. E. Eiben, and T. Vicsek, {\it Optimized Flocking of Autonomous Drones in Confined Environments}, Sci. Robot. {\bf 3},
eaat3536 (2018).

\bibitem{AraujoVolpe2023} N. A. M. Ara{\'u}jo, {\it et al.} {\it Steering self-organisation through confinement}, Soft Matter {\bf 19}, 1695 (2023).

\bibitem{Ebbens2016} S. J. Ebbens, {\it Active colloids: Progress and challenges towards realising autonomous applications}, Curr. Opin. Colloid Interface Sci. {\bf 21}, 14–23 (2016).

\bibitem{ShaebaniRieger2020} M. R. Shaebani, A. Wysocki, R. G. Winkler, G. Gompper, and H. Rieger, {\it Computational models for active matter} Nat. Rev. Phys. {\bf 2}, 181 (2020).

\bibitem{MognettiFrenkel2013} B. M. Mognetti, A. Saric, S. Angloletti-Uberti, A. Cacciuto, C. Valeriani, and D. Frenkel, {\it Living Clusters and Crystals from Low-Density Suspensions of Active Colloids}, Phys. Rev. Lett. {\bf 111}, 245702 (2013).
\bibitem{LeunissenBlaaderen2005} M. E. Leunissen, C. G. Gristova, A.-P. Hynninen, C. P. Royall, A. I. Campbell, A. Imhof, M. Dijkstra, R. van Roij, and A. van Blaaderen, {\it Ionic colloidal crystals of oppositely charged particles}, Nature (London) {\bf 437}, 235 (2005).

\bibitem{Klapp2016} S. H. L. Klapp, Curr. Opin. Colloid Interface Sci. {\bf 21}, 76 (2016).

\bibitem{ZiepkeFrey2022} A. Ziepke, I. Maryshev, I. S. Aranson, and E. Frey, {\it Multiscale organization in communicating active matter}, Nat. Commun. {\bf 13}, 6727 (2022).

\bibitem{BEC1}
M. H. Anderson, J. R. Ensher, M. R. Matthews, C. E. Wieman, and E. A. Cornell,
Observation of Bose-Einstein condensation in a dilute atomic vapor,
Science \textbf{269}, 198 (1995).

\bibitem{BEC2}
C. C. Bradley, C. A. Sackett, J. J. Tollett, and R. G. Hulet,
Bose-Einstein condensation of sodium atoms,
Phys. Rev. Lett. \textbf{75}, 1687 (1995).

\bibitem{leibfried2003quantum}
D. Leibfried, R. Blatt, C. Monroe, and D. Wineland,
\textit{Quantum dynamics of single trapped ions},
Rev. Mod. Phys. \textbf{75}, 281 (2003).


\bibitem{hu1994observation}
Z. Hu and H. J. Kimble,
\textit{Observation of a single atom in a magneto-optical trap},
Opt. Lett. \textbf{19}, 1888 (1994).

\bibitem{bloch2008many}
I. Bloch, J. Dalibard, and W. Zwerger, {\it Many-Body Physics with Ultracold Gases}, Rev. Mod. Phys. {\bf 80}, 885 (2008).


\bibitem{lewenstein2007ultracold}
M. Lewenstein, A. Sanpera, V. Ahufinger, B. Damski, A. Sen, and U. Sen,
\textit{Ultracold atomic gases in optical lattices: mimicking condensed matter physics and beyond},
Adv. Phys. \textbf{56}, 243 (2007).

\bibitem{langen2015ultracold}
T. Langen, R. Geiger, and J. Schmiedmayer, {\it Ultracold atoms out of equilibrium}, Annu. Rev. Condens. Matter Phys. {\bf 6}, 201 (2015).

\bibitem{polkovnikov2011colloquium}
A. Polkovnikov, K. Sengupta, A. Silva, and M. Vengalattore,
{\it Colloquium: Nonequilibrium dynamics of closed interacting quantum systems }, Rev. Mod. Phys. {\bf 83}, 863 (2011).

\bibitem{heyl2018dynamical}
M. Heyl,
\textit{Dynamical quantum phase transitions: a review},
Rep. Prog. Phys. \textbf{81}, 054001 (2018).


\bibitem{schneider2012fermionic}
U. Schneider, L. Hackermüller, J. P. Ronzheimer, S. Will, S. Braun, T. Best, I. Bloch, E. Demler, S. Mandt, D. Rasch, et al.,
\textit{Fermionic transport and out-of-equilibrium dynamics in a homogeneous Hubbard model with ultracold atoms},
Nat. Phys. \textbf{8}, 213 (2012).

\bibitem{Vicsek1995}
T. Vicsek, A. Czir{\'o}k, E. Ben-Jacob, I. Cohen, and O.
Shochet, {\it Novel Type of Phase Transition in a System of Self-Driven Particles}, Phys. Rev. Lett. {\bf 75}, 1226 (1995).

\bibitem{LopezClenebtt2015} H. M. L{\'o}pez, J. Gachelin, C. Douarche, H. Auradou, and E. Cl{\'e}ment, {\it Turning Bacteria Suspensions into Superfluids}, Phys. Rev. Lett. {\bf 115}, 028301 (2015).

\bibitem{GuoCheng2018} S. Guo, D. Samanta, Y. Peng, X. Xu, and X. Cheng, { \it Symmetric shear banding and swarming vortices in bacterial superfluids}, Proc. Natl. Acad. Sci. U.S.A. {\bf 115}, 7212 (2018).

\bibitem{teVrugtWittkowski2023}  M. te Vrugt, T. Frohoff-H{\"u}lsmann, E. Heifetz, U. Thiele, and R. Wittkowski, { \it From a microscopic inertial active matter model to the Schrödinger equation}, Nat. Commun. {\bf 14}, 1302 (2023).

\bibitem{SoneAshida2019} K. Sone and Y. Ashida, {\it Anomalous Topological Active Matter}, Phys. Rev. Lett. {\bf 123}, 205502 (2019).

\bibitem{LoeweGoldbart2018} B. Loewe, A. Souslov, and P. M. Goldbart, {\it Flocking from a Quantum Analogy: Spin–Orbit Coupling in an Active Fluid},
New J. Phys. {\bf 20}, 013020 (2018). 
\bibitem{CouderFort2006} Y. Couder, and E. Fort, {\it Single-Particle Diffraction and Interference at a Macroscopic Scale}, Phys. Rev. Lett. {\bf 97}, 154101 (2006).

\bibitem{AvronOaknin2006} J. E. Avron, B. Gutkin, and D. H. Oaknin, {\it Adiabatic Swimming in an Ideal Quantum Gas}, Phys. Rev. Lett. {\bf 96}, 130602 (2006).

\bibitem{Saito2015} H. Saito, {\it Can we swim in superfluids?: Numerical demonstration of self-propulsion in a Bose–Einstein condensate}, J. Phys. Soc. Jpn. {\bf 84}, 114001 (2015).

\bibitem{ShuklaPandit2016} V. Shukla, M. Brachet, and R. Pandit, {\it Sticking transition in a minimal model for the collisions of active particles in quantum fluids}, Phys. Rev. A {\bf 94}, 041602(R) (2016).

\bibitem{GiuriatoKrstulovic2019} U. Giuriato and G. Krstulovic, {\it Interaction between active particles and quantum vortices leading to Kelvin wave generation} Sci. Rep. {\bf 9}, 4839 (2019).

\bibitem{KolmakovAranson2021} G. V. Kolmakov and I. S. Aranson, {\it Superfluid swimmers}, Phys. Rev. Res. {\bf 3}, 013188 (2021).

\bibitem{AdachiKawaguchi2022} K. Adachi, K. Takasan, and K. Kawaguchi, {\it Activity-induced phase transition in a quantum many-body system} Phys. Rev. Res. {\bf 4}, 013194 (2022).

\bibitem{YamagishiObuse2023} M. Yamagishi, N. Hatano, and H. Obuse, {\it Defining a quantum active particle using non-Hermitian quantum walk}, Sci. Rep. \textbf{14}, 28648 (2024).
\bibitem{MatsoukasDelcampo2023} A. S. Matsoukas-Roubeas, F. Roccati, J. Cornelius, Z. Xu, A. Chenu, and A. del Campo,
{\it Non-Hermitian Hamiltonian deformations in quantum mechanics}, J. High Energy Phys. {\bf 1}, 1-31, 6 (2023).
\bibitem{quantumflocks}
R. Khasseh, S. Wald, R. Moessner, C. A. Weber, and M. Heyl,
\textit{Active quantum flocks},
\textit{arXiv:2308.01603}.


\bibitem{BlochNascimbene2012} Bloch, I., J. Dalibard, and S. Nascimb{\'e}ne, {\it Quantum simulations with ultracold quantum gases} Nat. Phys. {\bf 8}, 267 (2012).

\bibitem{Kaufman2021} A. M. Kaufman, K.-K. Ni, {\it Quantum science with optical tweezer arrays of ultracold atoms and molecules}, Nat. Phys. {\bf 17}, 1324 (2021).

\bibitem{Lahaye2020} A. Browaeys, T. Lahaye, {\it Many-body physics with individually controlled Rydberg atoms}, Nat. Phys. {\bf 16}, 132 (2020).

\bibitem{SchaferTakahashi2020}  F. Sch{\"a}fer, T. Fukuhara, S. Sugawa, Y. Takasu, and Y. Takahashi, {\it Tools for quantum simulation with ultracold atoms in optical lattices}, Nat. Rev. Phys. {\bf 2}, 411 (2020).

\bibitem{MorigiWineland2015} G. Morigi, J. Eschner, C. Cormick, Y. Lin, D. Leibfried, and D. J. Wineland, {\it Dissipative Quantum Control of a Spin
Chain}, Phys. Rev. Lett. {\bf 115}, 200502 (2015).


\bibitem{PesceVolpe2020} G. Pesce, P. H. Jones, O. M. Marag{\`o} and G. Volpe, {\it Optical tweezers: theory and practice}, Eur. Phys. J. Plus {\bf 135}, 949 (2020).

\bibitem{Ashkin1970} A. Ashkin, { \it Acceleration and trapping of particles by radiation pressure}, Phys. Rev. Lett. {\bf24}, 156–159 (1970)

\bibitem{ChenMuga2010} X. Chen, A. Ruschhaupt, S. Schmidt, A. del Campo, D. Guery-Odelin, and J. G. Muga, {\it Fast Optimal Frictionless Atom Cooling in Harmonic Traps: Shortcut to Adiabaticity}, Phys. Rev. Lett. {\bf 104}, 063002 (2010).

\bibitem{AbahLutz2012} O. Abah, J. Ro{\ss}nagel, G. Jacob, S. Deffner, F. Schmidt-Kaler, K. Singer, and E. Lutz, {\it Single-Ion Heat Engine at Maximum Power}, Phys. Rev. Lett. {\bf 109}, 203006 (2012).

\bibitem{RossnagelLutz2014} J. Ro{\ss}nagel, O. Abah, F. Schmidt-Kaler, K. Singer, and E. Lutz, {\it Nanoscale Heat Engine Beyond the Carnot Limit}, Phys. Rev. Lett. {\bf 112}, 030602 (2014).

\bibitem{SabhapanditMajumdar2024} S. Sabhapandit and S. N. Majumdar, {\it Noninteracting particles in a harmonic trap with a
stochastically driven center}, J. Phys. A:
Math. Theor. {\bf 57}, 335003 (2024).

\bibitem{UhlenbeckOrnstein1930} G. E. Uhlenbeck, and L. S. Ornstein, 1930, {\it On the theory of the Brownian motion} Phys. Rev. {\bf 36}, 823–841 (1930).

\bibitem{RomanczukGeier2012} P. Romanczuk, M. B{\"a}r, W. Ebeling, B. Lindner, and L. Schimansky-Geier, { \it Active Brownian particles}, Eur. Phys. J. Spec. Top. {\bf 202}, 1 (2012).
\bibitem{Bonilla2019} L. Bonilla, {\it Active Ornstein-Uhlenbeck particles}, Phys.~Rev.~E {\bf 100}, 022601 (2019).
\bibitem{RivasHuelga2011} \`A. Rivas and S. F. Huelga, {\it Open Quantum Systems: An Introduction} (Springer, Berlin, 2011).
\bibitem{EnglertMorigi2002} BG. Englert and G. Morigi. {\it Five Lectures on Dissipative Master Equations. Coherent Evolution in Noisy Environments}. Lecture Notes in Physics, vol 611. (Springer, Berlin, Heidelberg, 2002).
\bibitem{Lindbladcorr} A. P. Antonov, S. Lee, B. Liebchen, and H. L\"owen, J. Melles,
Y. Tuchkov, and M. te Vrugt (unpublished).
\bibitem{supplement} See Supplemental Material for details
of analytical calculations and numerical method,
which includes Refs.~\cite{Griffiths2018, GardinerZoller2000, BreuerPetruccione2002, Manzano2020, numerical, MartinvanWijland2021, combinatorics}

\bibitem{nguyen2021active} G.~H.~P. Nguyen, R. Wittmann, and H. L{\"o}wen, {\it Active Ornstein--Uhlenbeck model for self-propelled particles with inertia} J. Phys.: Condens. Matter \textbf{34}, 035101 (2021).

\bibitem{Veyron2024} R. Veyron, J.-P. G{\'e}rent, G. Baclet, V. Mancois, P. Bouyer, and S. Bernon, {\it In Situ Subwavelength Microscopy of Ultracold Atoms Using Dressed Excited States}, PRX Quantum \textbf{5}, 030349 (2024).

\bibitem{SolonCates2015} A. Solon, M. Cates, and J. Tailleur, {\it Active Brownian Particles and Run-and-Tumble Particles: a Comparative Study}, Eur. Phys. J. Spec. Top.
{\bf 224}, 1231 (2015).

\bibitem{CatesTailleur2013} M. E. Cates and J. Tailleur, {\it When are active Brownian particles and run-and-tumble particles equivalent? Consequences for motility-induced phase separation}, Europhys. Lett. {\bf 101}, 20 010 (2013).

\bibitem{FreyKroy2005}  E. Frey and K. Kroy, {\it Brownian motion: a paradigm of soft matter and biological physics}, Ann. Phys. (Leipzig) {\bf 517} (2005), pp. 20–50.

\bibitem{HyrkasManninen2013} M. Hyrk\"as, V. Apaja, and M. Manninen, {\it Many-particle dynamics of bosons and fermions in quasi-one-dimensional flat-band lattices}, Phys. Rev. A {\bf87}, 023614 (2013).

\bibitem{AbrahamBonitz2012} J. W. Abraham, K. Balzer, D. Hochstuhl, and M. Bonitz, {\it Quantum breathing mode of trapped systems in one and two dimensions}, Phys. Rev. B {\bf 86}, 125112 (2012).

\bibitem{ZhengPoletti2015} Y. Zheng and D. Poletti, {\it Quantum statistics and the performance of engine cycles}, Phys. Rev. E {\bf92} 012110 (2015).

\bibitem{MyersDeffner2020} N. M. Myers and S. Deffner, {\it Bosons outperform fermions: The thermodynamic advantage of symmetry}, Phys. Rev. E {\bf 101}, 012110 (2020).

\bibitem{GreinerBloch2002} M. Greiner, O. Mandel, T. Esslinger, T. W. H{\"a}nsch, and I. Bloch, {\it Quantum phase transition from a superfluid to a Mott insulator in a gas of ultracold atoms}, Nature (London) {\bf 415}, 39 (2002).
\bibitem{Griffiths2018} D. J. Griffiths and D. F. Schroeter, {\it Introduction to Quantum Mechanics}, third edition ed. (Cambridge University Press, Cambridge; New York, NY, 2018).
\bibitem{GardinerZoller2000} C.~W. Gardiner and P. Zoller. {\it Quantum Noise}. Springer, Berlin, (2000).
\bibitem{BreuerPetruccione2002} H.~P. Breuer and F. Petruccione. {\it The theory of open quantum systems}. Oxford University Press, (2002).
\bibitem{Manzano2020} D. Manzano, {\it A short introduction to the Lindblad master equation}{ AIP Advances} {\bf 10}, 2, 025106 (2020).
\bibitem{numerical}  W. H. Press, S. A. Teukolsky, W. T. Vetterling, and B. P. Flannery, Numerical Recipes, The Art of Scientific Computing, 3rd ed. (Cambridge University Press, Cambridge, 2007).
\bibitem{MartinvanWijland2021} D. Martin, J. O’Byrne, M. E. Cates, {\'E}. Fodor, C. Nardini, J. Tailleur, and F. van Wijland, {\it Statistical mechanics of active Ornstein-Uhlenbeck particles}, Phys. Rev. E {\bf 103}, 032607 (2021).
\bibitem{combinatorics} L. Comtet, {\it Advanced Combinatorics}. Springer Dordrecht, (1974).
\end{thebibliography}

\begin{thebibliography} {11}
	\bibitem{Griffiths2018_supp} D. J. Griffiths and D. F. Schroeter, {\it Introduction to Quantum Mechanics}, third edition ed. (Cambridge University Press, Cambridge ; New York, NY, 2018).
	
	\bibitem{Bonilla2019_supp} L. Bonilla, {\it Active Ornstein-Uhlenbeck particles}, Physical Review E {\bf 100}, 022601 (2019).
	
	\bibitem{comment} We note that this auto-correlation function is also valid for other active processes exhibiting similar Langevin-type dynamics, such as those with colored noise or non-Markovian noise correlations.
	
	\bibitem{RivasHuelga2011_supp} \`A. Rivas and S. F. Huelga, {\it Open Quantum Systems: An Introduction} (Springer, Berlin, 2011).
	
	\bibitem{GardinerZoller2000_supp} C.W. Gardiner and P. Zoller. {\it Quantum Noise}. Springer, Berlin, (2000).
	
	\bibitem{BreuerPetruccione2002_supp} H.P. Breuer and F. Petruccione. {\it The theory of open quantum systems}. Oxford University Press, (2002).
	
	\bibitem{Manzano2020_supp} D. Manzano, {\it A short introduction to the Lindblad master equation}{ AIP Advances} {\bf 10}, 2, 025106 (2020).
	
	\bibitem{numerical_supp}  W. H. Press, S. A. Teukolsky, W. T. Vetterling, and B. P. Flannery, Numerical Recipes, The Art of Scientific Computing, 3rd ed. (Cambridge University Press, Cambridge, 2007).
	
	\bibitem{MartinvanWijland2021_supp} D. Martin, J. O’Byrne, M. E. Cates, {\'E}. Fodor, C. Nardini, J. Tailleur, and F. van Wijland, {\it Statistical mechanics of active Ornstein-Uhlenbeck particles}, Phys. Rev. E {\bf 103}, 032607 (2021).
	
	\bibitem{nguyen2021active_supp} G.~H.~P. Nguyen, R. Wittmann, and H. L{\"o}wen, {\it Active Ornstein--Uhlenbeck model for self-propelled particles with inertia} J. Phys.: Condens. Matter \textbf{34}, 035101 (2021).
	\bibitem{combinatorics_supp} L. Comtet, {\it Advanced Combinatorics}. Springer Dordrecht, (1974).
\end{thebibliography}

\end{document}